\documentclass[11pt,psfig]{article}
\usepackage{apacite}

\usepackage{graphicx}
 \usepackage{amsmath}
 \usepackage{amssymb}
\usepackage{color}
\usepackage{placeins}

\usepackage{xcolor,colortbl}
\definecolor{Gray}{gray}{0.85}
\newcolumntype{g}{>{\columncolor{Gray}}c}
\newcolumntype{w}{>{\columncolor{white}}c}
\definecolor{LightCyan}{rgb}{0.88,1,1}

  \title{   Nonparametric estimation of the  
conditional distribution %with monotonicity correction
 at regression boundary points }

\author{Srinjoy Das \\
Department of Electrical \\ and Computer Engineering \\
     University of California---San Diego \\
   La Jolla, CA 92093, USA 
   \\email: {\tt   s2das@ucsd.edu}
\and
Dimitris N. Politis 
\\Department of Mathematics \\
     University of California---San Diego \\
   La Jolla, CA 92093-0112, USA 
  \\   email: {\tt   dpolitis@ucsd.edu} 
 }
\date{ } % version Nov. 2011

\begin{document}

\newcolumntype{g}{>{\columncolor{Gray}}c}
 \maketitle

\begin{abstract}
Nonparametric regression %via  kernel smoothing
 is a standard statistical tool with 
increased importance    in the Big Data era.
Boundary points    pose additional difficulties but local polynomial regression 
can be used to alleviate them.
Local linear regression, for example, is easy to implement and performs quite well
both at  interior as well as boundary points.   
Estimating the conditional distribution function and/or the quantile function
at a given regressor point 
is immediate via standard kernel methods but problems ensue if local linear methods are
to be used. In particular, the distribution function estimator is not guaranteed to be monotone
increasing, and the quantile curves can ``cross''. 
In the paper at hand, a simple method of correcting
the local linear distribution   estimator for monotonicity is proposed, and its
good performance is demonstrated    via simulations and real data examples. 

\end{abstract}

 
 \clearpage
 
%\footnote{add ref: 
%
%Yu, Keming, and M. C. Jones. "Local linear quantile regression." Journal of the American statistical Association 93, no. 441 (1998): 228-237.
%
%Yu, Keming, Zudi Lu, and Julian Stander. "Quantile regression: applications and current research areas." Journal of the Royal Statistical Society: Series D (The Statistician) 52, no. 3 (2003): 331-350.
%
%Koenker, Roger. Quantile regression. No. 38. Cambridge university press, 2005.
%
%Wand, M. P., and M. C. Jones.  Kernel Smoothing, Vol. 60 of Monographs on statistics and applied probability.  (1994) CRC Press. 
%
%}
%\footnote{add remark saying that the monotonicy correction can be applied to
%other estimators of the conditional distribution not neces. local linear}
%\footnote{add remark on asymp. equivalence of corrected to uncorrected. Conf Int via bootsttrap
%--see MF book}

 \section{Introduction}

Nonparametric regression via kernel smoothing is a standard statistical tool with 
increased importance    in the Big Data era; 
see e.g.  \cite{wand1994kernel}, \cite{yu1998local},  \cite{yu2003quantile}, \cite{koenker2005quantile} or  \cite{schucany2004kernel} for reviews. 
The fundamental nonparametric regression problem is estimating the regression function $  \mu(x ) = E(Y |X=x )$ from data
$(Y_1,   x_1), \ldots, (Y_n ,   x_n)$
under the sole assumption that the function $\mu (\cdot )$ belongs
to some smoothness class, e.g., that it possesses a certain number of
 continuous derivatives.
 Here,   $Y_i $ is the real-valued response  
associated with the regressor $X$ taking a value of $  x_i$. 
Either by design or via the conditioning, the regressor values  
$    x_1, \ldots,    x_n $ are treated as nonrandom.  
For simplicity of exposition, we will assume that 
  the  regressor $X$ is univariate but extension  to the multivariate case
is straightforward.

A common approach to nonparametric regression starts with assuming 
  that the data were generated by an additive model such as 
\begin{equation}
Y_i=\mu( x_i)+ \sigma (x_i) \epsilon_i  \ \ \mbox{for} \ \  i=1, 2, \ldots, n
 \label{MB_eq.model}
\end{equation}
where the errors $\epsilon_i$ are assumed to be independent, identically
distributed  (i.i.d.) with mean zero and variance one, and
 $\sigma (\cdot )$ is another unknown %smooth 
function.

Nevertheless, standard kernel smoothing methods are applicable in a Model-Free  
context as well, i.e., without assuming an equation such as (\ref{MB_eq.model}). 
An important example is the Nadaraya-Watson kernel estimator defined as 
\begin{equation} \label{MF3short.eq.NW}
	\hat \mu(x )= \frac{\sum_{i=1}^n    \tilde K_{i,x}Y_i}
{\sum_{i=1}^n    \tilde K_{i,x}}
\end{equation}
where $b>0$ is the bandwidth,  $K(x)$ is a nonnegative 
  kernel function satisfying  $\int K(x) dx=1$, and
$$ %\begin{equation} \label{MF3short.eq.NWweights}
\tilde K_{i,x}= \frac{1}{b} K\left(\frac{x-x_i}{b} \right) .
% \tilde K\left(\frac{x-x_i}{b} \right) =
% \frac{K\left(\frac{x-x_i}{b} \right)}{\sum_{k=1}^n K\left(\frac{x-x_k}{b} \right)} .
$$ %\end{equation}
%for simplicity, in what follows we will assume that $K$ is everywhere nonnegative.
Estimator $\hat \mu(x )$ enjoys favorable properties such as consistency
and asymptotic normality under standard regularity conditions in a Model-Free
context,  e.g. assuming the pairs $(Y_1,   X_1), \ldots, (Y_n ,   X_n)$
are i.i.d. \cite{li2007nonparametric}.
  
The rationale behind   the Nadaraya-Watson  estimator  (\ref{MF3short.eq.NW})
is approximating the unknown function $\mu (x)$ by a constant over a window
of  ``width'' $b$; this is made clearer if a rectangular function is chosen as
 the kernel $K$, e.g. letting $K(x)={\bf 1}\{|x|<1/2\}$ 
where ${\bf 1}_A$ is the indicator of set $A$, in which 
  case $\hat \mu(x )$ is just the average of the $Y$'s
whose $x$ value fell in the window.
 Going from a local constant to a local
linear approximation for  $\mu (x)$, i.e., a first-order Taylor expansion,
 motivates the local linear estimator
\begin{equation}
\hat \mu ^{LL}(x)=\frac{ \sum_{i=1}^{n } w_iY_i }{\sum_{i=1}^{n } w_i  }
 \label{NSTS.eq.locallinearF}
\end{equation}
where
\begin{equation} 
w_i= \tilde K_{i,x} \left(1-\hat \beta(x-x_i)\right)  
\ \ \mbox{and} \ \ 
\hat \beta =\frac{\sum_{i=1}^n\ \tilde K_{i,x}(x-x_{i})}{ \sum_{i=1}^n\ \tilde K_{i,x}(x-x_{i})^2 }.
\label{NSTS.eq.locallinearweights}
 \end{equation}

If the  design points $x_j$ are  (approximately)  uniformly distributed 
over an interval $[a_1, a_2]$, then   $\hat \mu ^{LL}(x)$
 is typically indistinguishable from the   Nadaraya-Watson  estimator 
$\hat \mu  (x)$ when $x$ is in the `interior', i.e., when $x\in  [a_1+b/2, a_2-b/2]$.
The local linear estimator   $\hat \mu ^{LL}(x)$ offers an advantage
when the  design points $x_j$ are    non-uniformly distributed, e.g., when there
are gaps in the design points, and/or when $x$ is a {\it boundary} point, i.e., when $x=a_1 $
or $ x=a_2 $ (plus or minus $b/2$); see \cite{fan1996local} for details.
 
%\footnote{I replaced the notation $D_x(y)$ to the more standard ${F}_x(y)$
%(I used ${F}$ so we can find those changes easily)}

Instead of focusing on  the conditional moment $  \mu(x ) = E(Y |X=x )$, one may 
consider estimating the   conditional  distribution 
$ {F}_{x}(y) = P(Y  \leq y | X  = x) $ at some fixed point~$y$.
Note that  $ {F}_{x}(y) =E (W |X=x )$ where $W= {\bf 1}\{ Y  \leq y\}  $.
Hence, estimating $ {F}_{x}(y)$ can be easily done via local constant or
local linear estimation of the conditional mean from the new
dataset $(W_1,   x_1) \ldots, (W_n ,   x_n)$ where $W_i= {\bf 1}\{ Y_i  \leq y\}  $.
To elaborate, the local constant and the 
local linear estimators of $ {F}_{x}(y)$ are respectively given by  
 
\begin{equation} \label{MF.eq.hatD}
	\hat {F}_{x}(y) = \frac{\sum_{i=1}^n \tilde K_{i,x} {\bf 1}\{ Y_i  \leq y\}    } 
{\sum_{i=1}^n    \tilde K_{i,x}}  , 
\ \ \mbox{and} \ \ 
 \hat {F}_{x}^{LL}(y)=\frac{ \sum_{j=1}^{n } w_j{\bf 1}\{ Y_j  \leq y\}  }{\sum_{j=1}^{n } w_j  }
 \end{equation}
where the local linear weights $w_j$ are given by eq.~(\ref{NSTS.eq.locallinearweights}).

Viewed as a function of $y$,  
$\hat {F}_{x}(y)$ is a well-defined distribution function; however, being a local
constant estimator, it often has poor performance at boundary points. As
already discussed,  $\hat {F}_{x}^{LL}(y)$ has better performance at boundary points.
  Unfortunately, $\hat {F}_{x}^{LL}(y)$ is neither guaranteed to be in $[0,1]$
nor is it   guaranteed to be nondecreasing as a function of $y$; this is due to some of 
the weights $w_j$ potentially being negative.   

The problem  with non-monotonicity of $\hat {F}_{x}^{LL}(y)$  and the associated 
  quantile curves  potentially  ``crossing'' %in quantile regression 
is well-known;  
 see \cite{Hall1999} for the former issue, and \cite{yu1998local} for the latter, as well as the reviews on quantile regression by 
\cite{yu2003quantile} and \cite{koenker2005quantile}. In the next section, a simple method of correcting
the local linear distribution   estimator for monotonicity is proposed; its
good  performance is demonstrated   via simulations and real data examples
in Section 3. It should be noted here that while the paper at hand focuses on the monotonicity correction
for local linear estimators of the conditional distribution,  the monotonicity correction idea 
 can equally be 
be applied to other distribution estimators constructed via  different nonparametric methods, e.g. wavelets.

 \section {Local Linear Estimation of smooth conditional distributions}
 \label{LL_est}

\subsection{Some issues with current methods} 
The good performance of local constant and  local linear estimators 
({\ref{MF.eq.hatD}}) hinges on    
$ {F}_{x}(\cdot) $ being smooth, e.g. continuous, as a  function  of $x$.
In all that follows, we will further assume that $ {F}_{x}(y) $ is also continuous  in 
$y$ for all $x$. Since the estimators 
({\ref{MF.eq.hatD}}) are discontinuous (step functions) in $y$, it is customary
to smooth them, i.e., define
\begin{equation} \label{MF.eq.barD}
\bar {F}_{x}(y) = \frac{\sum_{i=1}^n \tilde K_{i,x} \Lambda(\frac {y-Y_{ {i}}} {{h}_0})}  
{\sum_{i=1}^n \tilde K_{i,x}  }  , 
\ \ \mbox{and} \ \ 
 \bar {F}_{x}^{LL}(y)=\frac{ \sum_{j=1}^{n } w_j\Lambda(\frac {y-Y_{ {j}}} {{h}_0})  }{\sum_{j=1}^{n } w_j  }
 \end{equation}
where $\Lambda(y)$ is some smooth distribution function which is strictly increasing 
with density   $\lambda(y)>0$, i.e., $\Lambda(y)=\int_{-\infty}^y \lambda(s)ds  $. 
Here again the local linear weights $w_j$ are given by eq.~(\ref{NSTS.eq.locallinearweights}),
and ${h}_0>0$ is a secondary bandwidth whose choice is not as important as the choice
of $b$; see Section \ref{MF_bw} for 
 %see \cite{li2007nonparametric} and Politis (\footnote{BOOK}) for
some concrete suggestions on picking  $b$ and   ${h}_0$ in practice. 

Under standard conditions, both estimators appearing in eq.~(\ref{MF.eq.barD}) 
are asymptotically consistent, and preferable to the respective estimators appearing in eq.~(\ref{MF.eq.hatD}), i.e., replacing ${\bf 1}\{ Y_j  \leq y\}   $
by $\Lambda(\frac {y-Y_{ {j}}} {{h}_0})$ is advantageous; see Ch.~6 of  \cite{li2007nonparametric}.
 Furthermore, as discussed in the Introduction, we expect that 
$\bar {F}_{x}^{LL}(y)$ would be a better estimator than $\bar {F}_{x}(y)$ when 
$x$ is a boundary point and/or the design is not uniform,
 while $\bar {F}_{x}^{LL}(y)$ and $\bar {F}_{x}(y)$ 
would be practically equivalent when $x$ is an interior point and the design
is (approximately) uniform. Hence, all in all, $\bar {F}_{x}^{LL}(y)$ 
would be preferable to  $\bar {F}_{x}(y)$
as an estimator of $  {F}_{x}(y)$ for   any fixed $y$. 
  The problem again is that
 $\bar {F}_{x}^{LL}(y)$ is not guaranteed to be a proper distribution 
viewed as a function of $y$   by analogy to $\hat {F}_{x}^{LL}(y)$.

There have been several proposals in the literature to address this issue.
%\footnote{also mention the method of Chernozukhov et al.
%for estimation of ${F}_x$ but say it is too cumbersome, etc}
An interesting one   is the adjusted Nadaraya-Watson estimator of 
\cite{Hall1999} that is a linear function of the $Y$'s with weights
being selected by an appropriate optimization procedure.  The adjusted Nadaraya-Watson estimator is much like a local linear estimator in that it has reduced bias 
(by an order of magnitude) compared to the regular Nadaraya-Watson local constant estimator. Unfortunately,
the adjusted Nadaraya-Watson estimator does not work well when $x$ is a boundary point
as the required optimization procedure typically  does not admit a solution.

Noting that the problems with $\bar {F}_{x}^{LL}(y)$ and $\hat {F}_{x}^{LL}(y)$
arise due to potentially negative  weights $w_j$ computed by eq.~(\ref{NSTS.eq.locallinearweights}), Hansen proposed
  a straightforward adjustment to  the local linear estimator    
that maintains its favorable asymptotic
properties \cite{hansen2004nonparametric} . The local linear  versions of $\hat {F}_{x}(y) $ and $\bar {F}_{x}(y) $ adjusted via Hansen's  proposal 
% \cite{hansen2004nonparametric} 
 are given as follows:
\begin{equation}
\label{MF.eq.sll_cdf}
\hat {F}_{x}^{LLH}(y) = \frac{\sum_{i=1}^{n} w_{i}^\diamond {\bf 1}(Y_{ i} \le y)}{\sum_{i=1}^{n} w_{i}^\diamond}  
\ \ \mbox{and} \ \ 
\bar {F}_{x_m}^{LLH}(y) = \frac{\sum_{i=1}^{n} w_{i}^\diamond 
\Lambda(\frac {y-Y_{ i}} {{h}_0})}{\sum_{i=1}^{n} w_{i}^\diamond}  
\end{equation}
where 
\begin{align}
\label{MF.eq.sll_diamond}
w_{i}^\diamond = \begin{cases}
                 \ 0  & \ \mbox{when} \ \hat \beta(x-x_{i}) > 1 \\
                 \ \tilde K_{i,x} \left(1 - \hat \beta(x-x_{i})\right) & \ \mbox{when} \ \hat \beta(x-x_{i}) \le 1 .
                 \end{cases}
\end{align}

\noindent Essentially, Hansen's  %\cite{hansen2004nonparametric} 
proposal replaces  negative weights by zeros, and then renormalizes the nonzero
weights. The problem here is that if $x$ is on the boundary, negative
weights are crucially needed in order to ensure an extrapolation takes place
with minimal bias; this is further elaborated upon in the following subsection. 

\subsection{Extrapolation vs.~interpolation} 

In order to illustrate the need for negative weights consider 
the simple case of $n=2$, i.e., two data points  $(Y_1,   x_1) $ and $ (Y_2 ,   x_2)$.
The question is to predict a future response $Y_3$ 
associated with a regressor value of $x_3$; assuming finite second moments,
the $L_2$--optimal predictor of  $Y_3$  is $\mu (x_3)$ where $
\mu(x ) = E(Y |X=x )$ as before. 

 If $x_3$ is an interior point as depicted in Figure  \ref{mf_internal}, the problem is
one of {\it interpolation}. If $x_3$ is a boundary point, and in particular 
if $x_3$ is outside the convex hull of the design points
 as in Figure  \ref{mf_boundary},  the problem is one of {\it extrapolation}.  
Let $\hat \mu ^{LL} (x) $ denote the local linear estimator of $\mu(x )$  as before. 
With  $n=2$, $\hat \mu ^{LL} (x) $  reduces to just
finding the line that passes through the  two data points  $(Y_1,   x_1) $ and
 $ (Y_2 ,   x_2)$. 
In other words, $\hat \mu ^{LL} (x) $ reduces to a convex combination of $Y_1$ and $Y_2$,
i.e., $\hat \mu ^{LL} (x) = \omega_x  Y_1+ (1-\omega_x)Y_2$ 
where $\omega_x= \frac{x_{2}-x}{x_{2}-x_{1}}$ where $x_1 < x < x_2$ for interior points
and $x_1 < x_2 < x$ for boundary points.
%\footnote{Srinjoy,  is the equation $\omega_x= \frac{x_{2}-x}{x_{2}-x_{1}}$ 
% the same even when $x>x_2$? If yes, please state clearly.}
Note that $\omega_x \in [0,1]$ if 
$x$ is an interior point, whereas $\omega_x \not \in [0,1]$ if 
$x$ is outside the convex hull of the design points. Hence, negative weights are
a {\it sine qua non} for effective linear extrapolation.
  
For example, assume we are in the setup of Figure  \ref{mf_boundary}  
where $x_1<x_2<x_3$. In this case, $\omega_{x_3}$ is negative. 
Hansen's proposal \cite{hansen2004nonparametric} 
 would  replace $\omega_{x_3}$ by zero and renormalize the coefficients,
 leading to $\hat \mu ^{LLH} (x_3) =  Y_2$; it is apparent that
this does not give the desired linear extrapolation effect.

To generalize the above setup, suppose that now $n$ is an arbitrary even number, and     $Y_i$   
represents the average  of   $n/2$ responses associated with regressor
value $x_i$ for $i=1$ or 2. Thus, we have a {\it bona fide} $n$--dimensional
scatterplot that is supported on two design points. Interestingly,  the
formula for $\hat \mu ^{LL} (x) $ is exactly as given above, and so is the
argument requiring negative weights for linear extrapolation.
 Of course, we cannot expect a general scatterplot to
be supported on just two design points. 
Nonetheless, in a nonparametric situation  one performs a linear regression {\it locally},
i.e., using a local subset of the data. Typically, there is a scarcity of
design points near the boundary, and the  general situation is 
qualitatively similar to the case of two design points.

\graphicspath{{/Users/rumpagiri/Documents/NONPARAMETRIC/model_free/papers/REGRESSION}}
\DeclareGraphicsExtensions{.png}

{\begin{figure}[!t]
  \centering
  \includegraphics[width=3.5in, height=2.5in]{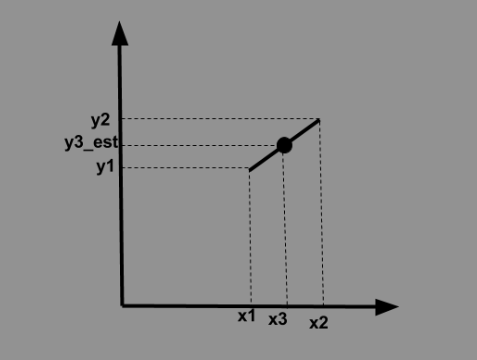}
  \caption{Interpolation: prediction of $Y_3$ when $x_3$ is an interior point;
$\hat Y_3$ is a convex combination of $Y_1$ and $Y_2$ with nonnegative weights.}
  \label{mf_internal}
\end{figure}}

{\begin{figure}[!t]
  \centering
  \includegraphics[width=3.5in, height=2.5in]{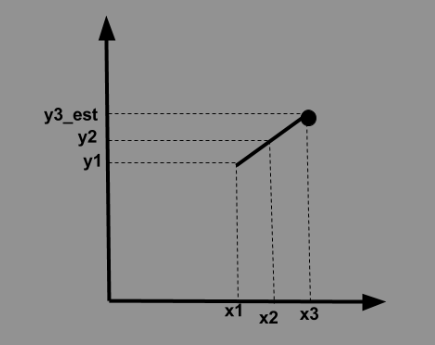}
  \caption{Extrapolation: prediction of $Y_3$ when $x_3$ is outside the convex hull of the design points;
$\hat Y_3$ is a linear combination of $Y_1$ and $Y_2$ with one positive and one
 negative weight.}
  \label{mf_boundary}
\end{figure}}

\clearpage

\subsection {Monotone Local Linear Distribution  Estimation}

%Looking at the form of estimator $\hat {F}_{x}^{LL }(y) $ from eq.~({\ref{MF.eq.hatD}}), it is
%apparent that the only way to guarantee that it is monotone as a function of $y$
%is to ensure that the weights $w_j$ are nonnegative; this is exactly 
%the   proposal that leads to estimator $\hat {F}_{x}^{LLH}(y) $ \cite{hansen2004nonparametric}.

%However, things are different when we consider  estimator $\bar {F}_{x}^{LL }(y) $ 
%from eq.~({\ref{MF.eq.barD}}); since $\bar {F}_{x}^{LL }(y) $ 
%  is, in any case,  preferable to $\hat {F}_{x}^{LL }(y) $,  we focus our
%attention on $\bar {F}_{x}^{LL }(y) $ from here on.

The estimator $\hat {F}_{x}^{LL }(y) $ from eq.~({\ref{MF.eq.hatD}}) is discontinuous as a function
of $y$ therefore we will focus our attention on
%the double-smoothed estimator 
$\bar {F}_{x}^{LL }(y) $ described in eq.~({\ref{MF.eq.barD}}) from here on. It seems that with this
double-smoothed estimator $\bar {F}_{x}^{LL }(y) $ we can ``have our cake and eat it too'', 
i.e., modify it towards monotonicity while  retaining (some of) the negative weights that are helpful
in the extrapolation problem as discussed in the last subsection. We are thus led to define a new
estimator denoted by $\bar {F}_{x}^{LLM }(y) $ which is a monotone version of $\bar {F}_{x}^{LL }(y) $;
we will refer to $\bar {F}_{x}^{LLM }(y) $  as the {\it Monotone Local Linear Distribution Estimator}.  

One way to define $\bar {F}_{x}^{LLM }(y) $ is as follows.
\\

\noindent {\bf Algorithm 1.}
\begin{enumerate} 
\item Compute  $\bar {F}_{x}^{LL }(y) $,  and denote $l=\lim_{y\to - \infty} \bar {F}_{x}^{LL }(y) $. 
\item Define a function $G_1(y)=  \bar {F}_{x}^{LL }(y)-l$. 
\item  Define a second function $G_2$ with the property that $G_2(y+\epsilon)=  \max 
\left( G_1(y+\epsilon), G_1(y) \right)$ for all $y$ and all $\epsilon >0$.
\item Define $\bar {F}_{x}^{LLM }(y) =G_2(y)/L$ where   $L=\lim_{y\to   \infty} G_2(y) $. 
\end{enumerate}
The above algorithm could be approximately implemented in practice by
selecting a small enough 
$\epsilon >0$,    dividing the
range of the $y$ variable using a grid of size $\epsilon  $, and running step~3 of Algorithm 1 sequentially. To elaborate, one would compute $G_2$ at grid point $j+1$ from the values of $G_1$ at previous, i.e., smaller, grid points. \\

A different---albeit equivalent---way of constructing the estimator
 $\bar {F}_{x}^{LLM }(y) $ is
by first constructing its associated density function denoted by $\bar f_{x}^{LLM }(y) $
which  will be called the {\it Monotone Local Linear
Density Estimator}. 
The alternative algorithm goes as follows.
\\

\noindent {\bf Algorithm 2.}
\begin{enumerate}
\item Recall that the derivative of $\bar {F}_{x}^{LL  }(y) $ 
with respect to $y$ is given by 
$$  \bar f_{x}^{LL}(y)=\frac{\frac {1} {{h}_0} \sum_{j=1}^{n } w_j  \lambda(\frac {y-Y_{ {j}}} {{h}_0})  }{\sum_{j=1}^{n } w_j  }  
 $$
where $\lambda(y)$ is the derivative of  $\Lambda(y)$.
\item Define a nonnegative version of $  \bar f_{x}^{LL}(y)$ as
$  \bar f_{x}^{LL+}(y)=\max (\bar f_{x}^{LL}(y), 0)$.

\item To make the above a proper density function, renormalize
it to area one, i.e., let 
\begin{equation}
\label{eq.densityMLL}
\bar f_{x}^{LLM }(y) = \frac{\bar f_{x}^{LL+}(y)  } {\int_{-\infty}^\infty \bar f_{x}^{LL+}(s)ds  }.
  \end{equation}
   
\item Finally, define $\bar {F}_{x}^{LLM }(y) =\int_{-\infty}^y \bar f_{x}^{LLM }(s) ds.$
\end{enumerate}

\noindent 
To implement the above one would again need to divide the
range of the $y$ variable using a grid of size $\epsilon  $ in order to
construct the maximum function in step~2 of Algorithm 2.  The same grid
can by used to provide Riemann-sum approximations to the two integrals
appearing in steps~3 and 4. All in all, the implementation of 
Algorithm 1 is a bit easier, and will be employed in the sequel. 
 
\subsection{Standard Error of the Monotone Local Linear
  Estimator}

 Under standard conditions, the local linear estimator 
$\sqrt{nb} \bar {F}_{x}^{LL }(y)$ is asymptotically normal with a
 variance $V_{x,y}^2$ that depends on the design; for details, see Ch.~6 of
\cite{li2007nonparametric}. 
In addition, the bias of $\sqrt{nb} \bar {F}_{x}^{LL }(y)$ is asymptotically vanishing 
if $b=o(n^{1/5})$. Hence, letting $b\sim n^\alpha$ for some $\alpha \in (0,1/5)$,
$  \bar {F}_{x}^{LL }(y)$ will be  consistent for $    {F}_{x} (y)$, and approximate 
95\% confidence intervals for $    {F}_{x} (y)$ can be 
constructed as  $\bar {F}_{x}^{LL }(y)\pm 1.96 \frac{V_{x,y}}{nb} $. 

The  consistency of $  \bar {F}_{x}^{LL }(\cdot)$ towards $    {F}_{x} (\cdot)$
implies that the monotonicity corrections described in the previous
subsection will be asymptotically negligible. 
To see why, fix a point $x$ of interest, and
assume that $    {F}_{x} (y)$ is absolutely continuous with 
density $    {f}_{x} (y)$ that is strictly positive over its support. 
The consistency of $\bar f_{x}^{LL}(y)$ to a positive target implies that
$\bar f_{x}^{LL}(y)$ will eventually become (and stay) positive
as $n$ increases. 
Hence, the monotonicity correction eventually vanishes, and $  \bar {F}_{x}^{LLM }(y)$
is asymptotically equivalent  to $  \bar {F}_{x}^{LL }(y)$.

Regardless, it is not advisable to use the aforementioned asymptotic distribution and variance 
of $  \bar {F}_{x}^{LL }(y)$ to approximate those of $  \bar {F}_{x}^{LLM }(y)$ for practical work
since, in finite samples, $  \bar {F}_{x}^{LLM }(y)$ 
and  $  \bar {F}_{x}^{LL }(y)$ can be quite different. 
Our recommendation is to use some form of bootstrap in order to approximate the 
distribution and/or standard error of  $  \bar {F}_{x}^{LLM }(y)$ directly. 
In particular, the Model-Free bootstrap   \cite{politis2015model}  in its many forms
is immediately applicable in the present context. For instance, the ``Limit Model-Free'' (LMF) 
bootstrap   would go as follows:
\\

\noindent {\bf LMF Bootstrap Algorithm}  %\cite{politis2015model}.}
\begin{enumerate}
\item Generate $U_1,\ldots, U_n$ i.i.d.~Uniform(0,1).
\item Define $Y_i^*= G^{-1}_{x_i}(U_i)$ for $i=1,\ldots, n$ 
where $ G^{-1}_{x_i}(\cdot)$ is the quantile inverse of 
$ \bar {F}_{x_i}^{LLM}(\cdot)$, i.e., $ G^{-1}_{x_i}(u)=\inf \{y: \bar {F}_{x_i}^{LLM}(y)\geq 
   u\}$. 
\item For the points $x$ and $y$ of interest, construct the pseudo-statistic 
  $\bar {F}_{x}^{LLM* }(y) $ which is computed by applying 
 estimator $\bar {F}_{x}^{LLM  }(y) $ to the bootstrap dataset
$(Y_1^*,   x_1) \ldots, $ $(Y_n^* ,   x_n)$.
\item Repeat steps 1--3 a large number (say $B$) times. Plot the 
$B$ pseudo-replicates $\bar {F}_{x}^{LLM* }(y) $ in a histogram that will serve as an 
approximation of the distribution of $\bar {F}_{x}^{LLM  }(y) $. 
In addition, the sample variance
of the $B$ pseudo-replicates $\bar {F}_{x}^{LLM* }(y) $ is the bootstrap estimator 
of the variance of $\bar {F}_{x}^{LLM  }(y) $.
\end{enumerate}
\noindent Our focus is on point estimation of $  {F}_{x} (y)$
so we will not elaborate further on the construction of
interval estimates. 
  
\subsection {Bandwidth Choice}
\label{MF_bw}

There are two bandwidths, $b$ and $h_0$, required to construct estimator $\bar {F}_{x}^{LLM  }(y) $
and its relatives $\bar {F}_{x}(y)$ and $\bar {F}^{LLH}_{x}(y)$.    We will now focus on
choice of the bandwidth $b$ which 
is the most crucial of the two, and  is often picked via leave-one-out cross-validation. 

In the paper at hand we are mostly concerned with  estimation and prediction at boundary points.
Since often boundary problems present their own peculiarities, we are strongly recommending 
carrying out the cross-validation procedure `locally', i.e., over a neighborhood of the point of interest. One
needs, however, to ensure that there are enough points nearby to perform the leave-one-out experiment.
Hence, our concrete recommendation goes 
  as follows. 
\begin{itemize}
 
\item Choose a positive integer $m$ which can be fixed number or it can be 
a small fraction of the sample size at hand. 

\item Then, identify   $m$ among the 
regression points $(Y_1, x_1), \ldots,(Y_n, x_n)$  with the property that their respective $x_i$'s are the
$m$ closest neighbors of the point $x$ under consideration.

\item Denote this set of $m$ points by $(Y_{g(1)} , x_{g(1)}), \ldots,(Y_{g(m)}, x_{g(m)})$ where the function $g(\cdot)$ gives the index numbers of the selected points. 

\item For $k=1,\ldots, m$, compute $\hat Y_{g(k)} $ which is the $L_2$--optimal predictor 
of $Y_{g(k)}$ using leave-one-out data. In other words,  $\hat Y_{g(k)} $ is the mean, 
i.e., center of location, of  one of the aforementioned distribution estimators
%$\bar {F}_{x}(y)$, $\bar {F}^{LLH}_{x}(y)$ or $\bar {F}^{LLM}_{x}(y)$. This is done 
based on the delete--one dataset, i.e. pretending that  $Y_{g(k)}$ is unavailable.

\item Thus, for a range of values of bandwidth $b$, we can calculate the following:

\begin{equation}
Err = \sum_{k=1}^{m} (\hat Y_{g(k)} - Y_{g(k)})^2
\label{MF.cv_ts_b_1}. 
\end{equation}
\item 
Finally, the optimal bandwidth is given by the value of $b$ that minimizes 
$Err$ over the range of   bandwidths considered.
\end{itemize}

% of $ \mu (x_{k+1})= E(Y_{k+1}|X_{k+1}=x_{k+1})$ using one of the methods described in the
% Introduction with bandwidth equal to $b$. 

%\footnote{Srinjoy this needs more elaboration: how do you get
%the estimated value of $\mu (x_{k+1})$ ? what  method are you using? what data 
%are you leaving out in order to do cross-validation? Can you do regular (two-sided)
%cross-validation as well here?} 
% Estimation of ${F}_x(y)$ is done
% using equations (\ref{MF.eq.barD}), (\ref{MF.eq.sll_cdf}) or (\ref{MF.sll_cdf_nonmono}) for local constant, local linear and monotone local linear
% density estimation respectively . 
%It is also possible to find $b_{cv}$ using an L1 criterion as below:

%\begin{equation}
%Err_{L1}(b) = \sum_{k=k_0}^{k=n} |\hat Y_{k+1}(b) - Y_{k+1}| 
%\label{MF.cv_b_1}
%\end{equation}

% For a single dataset cross-validation as described above is an effective way to find the optimal bandwidth $b$ to construct the best estimator
% for $\hat Y_{n+1}$. When $N$ realizations of  generated data $(Y_1, x_1) \ldots, (Y_{n+1} , x_{n+1})$ are available it is also possible to find the optimal
%bandwidth $b$ which gives the lowest average error as per an L2 or L1 criterion across all $N$ realizations:
% \begin{equation}
% \sum_{i=1}^{i=N}(\hat Y_{i,n+1}(b) - Y_{i,n+1})^2 \ \ \\\mbox {or}  \sum_{i=1}^{i=N} |\hat Y_{i,n+1}(b) - Y_{i,n+1}|
% \label{MF.synth_b}
 %\end{equation}

Coming back to the problem of selecting $h_0$, define   $h=b/n$ and recall that in an analogous regression problem  the optimal rates
$h_0\sim n^{-2/5}$ and $h \sim n^{-1/5}$  were suggested in connection with the nonnegative kernel $K$;
see  \cite{li2007nonparametric}. As in  \cite{Politis2013},  this leads to the practical
recommendation of letting $h_0=h^2$. We will adopt the same rule-of-thumb here as well,
namely let ${h}_0=b^2/n^2$ where $b$ has been chosen previously via local  cross-validation.
Note that the initial choice of $h_0$ (before performing the cross-validation to determine the optimal bandwidth $b$) 
can be set by a plug-in rule as available in standard statistical software such as R.

%\footnote{in order to choose $b$ we use cross-validation
%but what  $h_0$ are we using  while trying to choose $b$? I would guess
%use $h_0$ very small (practically zero) at that stage}

\section {Numerical work: simulations and real data}
\label{MF_sims}

%\footnote{Section needs complete rewriting using the language of the first part (not Model-free
%vs model-based)}

The performance of the three distribution estimators $\bar {F}_{x}(y)$, $\bar {F}^{LLH}_{x}(y)$ or $\bar {F}^{LLM}_{x}(y)$ 
described above are empirically compared using both simulated and real-life datasets
according to the following metrics. 
% These 3 estimators are referenced using the suffixes LC, LLH and LLM respectively in all the performance tables and plots shown hereafter.
% Following metrics are used for measuring and comparing the performance of the 3 estimators:
\begin{enumerate}
\item Divergence between the local distribution $\bar {F_x}(\cdot)$
%\footnote{do you mean $F_x (\cdot)$?} 
estimated by all three methods 
% at specified boundary and internal points  
and the corresponding local (empirical) distribution calculated from the actual data; this is determined using
  the mean value of the Kolmogorov-Smirnov (KS) test statistic. The measurement is performed on simulated datasets
where multiple realizations of data at both boundary and internal points are available. Therefore the empirical distribution at any
point can be calculated and compared versus the estimated values. Our notation  
is {\bf KS-LC, KS-LLH and KS-LLM} for the
distribution estimators $\bar {F}_{x}(y)$, $\bar {F}^{LLH}_{x}(y)$ or $\bar {F}^{LLM}_{x}(y)$ respectively.
\item Comparison of estimated quantiles of $F_x (\cdot)$ at specified points using all three methods versus the corresponding empirical values calculated using simulated datasets.
\item Point prediction performance as indicated by bias and Mean Squared Error (MSE) on simulated and real-life datasets using all three methods.
%It is possible to perform point prediction using fitted or predictive residuals for each of the three  distribution estimators \cite{Politis2013}. However the results are very similar
%for these 2 cases on all the datasets studied as expected. Therefore in this paper all point prediction results use fitted residuals
The MSE values of point prediction
are denoted as {\bf MSE-LC, MSE-LLH and MSE-LLM} for the distribution estimators $\bar {F}_{x}(y)$, $\bar {F}^{LLH}_{x}(y)$ or $\bar {F}^{LLM}_{x}(y)$ respectively; the 
corresponding bias values are denoted {\bf Bias-LC, Bias-LLH and Bias-LLM}.
% Point Prediction performance using Model-Based local linear estimation as given by (\ref{MB_mu}) is also shown for reference. 
For comparison purposes the point-prediction performance is also measured using the local linear conditional moment estimator
as given by equations \ref{NSTS.eq.locallinearF} and \ref{NSTS.eq.locallinearweights}. In this case bias and MSE are indicated
as {\bf Bias-LL} and {\bf MSE-LL} respectively.

% Note that in the Model-Free case performance is estimated using both the fitted and predictive approaches whereas in the Model-Based case no such distinction is made
% as the standard local linear approach is used for prediction.

\end{enumerate}

On simulated datasets the performance metrics for all three distribution estimators are calculated both at boundary and internal points
to illustrate how performance varies between $\bar {F}_{x}(y)$, $\bar {F}^{LLH}_{x}(y)$ and $\bar {F}^{LLM}_{x}(y)$  in the two cases.
 
\subsection{Simulation: Additive model with i.i.d Gaussian errors}

% \footnote{rewrite in a descriptive way: Data $Y_i$  were simulated via model BLA BLA...
% you must produce a readable text!}

Data $Y_i$ for $i =1, \ldots, 1001$ were simulated as per model (\ref{MB_eq.model})
by setting $\mu (x_i)= \sin(x_i)$, $\sigma(x_i)=\tau$ and the errors
$\epsilon_i$ as i.i.d.~$N(0,1)$. 
Sample size $n$ was set to 1001. 
%\footnote{why 1001 and not 1000}
A total of 500 such realizations were generated for this study. \\
% Point Prediction performance including bias and mean-square error (MSE) are studied for the 3 MF estimators - local constant (LC), local linear(LL)
% and monotone local linear(LLM) using a range of bandwidths - $10, 20, \ldots, 140$. 
\indent Results for the mean-value of the Kolmogorov-Smirnov test statistic between the LC, LLH and LLM estimated distributions and  empirical distribution calculated using available values of the simulated data
% \footnote{what do you mean "empirical distribution "?}
 are given in Tables \ref{DE_pp_iid_bp_1}, \ref{DE_pp_iid_ip_1}, \ref{DE_pp_iid_bp_2} and \ref{DE_pp_iid_ip_2} for boundary point $n=1001$
and internal point $n=200$ for values of $\tau=0.1$ and $0.3$ over a range  of bandwidths,
i.e., $b$ taking values $10, 20, \ldots, 140$. 
% \footnote{spell out KS; explain "Quantile plots "}

Point prediction performance values are provided for the same cases in Tables \ref{MF_pp_iid_bp_1}, \ref{MF_pp_iid_ip_1}, \ref{MF_pp_iid_bp_2} and \ref{MF_pp_iid_ip_2}.

\indent Estimates of the $\alpha$--quantile  at specific values of $\alpha$ are calculated using all three distribution estimators and compared with corresponding
% \footnote{how do you get the true?} 
quantiles calculated from the available data. Plots for selected quantile values ($\alpha = 0.1$ and $\alpha = 0.9$) are shown in Figures \ref{figure_quantile_01_1S},
\ref{figure_quantile_09_1S}, \ref{figure_quantile_01_2S} and \ref{figure_quantile_09_2S} for both 1 and 2-sided cases ($\tau=0.3$). Note that the 'true' quantile
lines showed in the plots are values calculated from the available data at $n = 1001$ and $n = 200$ over 500 realizations for the case of boundary and internal
points respectively. The bandwidths used for estimating the quantiles for LC, LLH and LLM are based on bandwidth values where the best performance for these estimators was
obtained using the Kolmogorov-Smirnov test (refer Tables \ref{DE_pp_iid_bp_2} and \ref{DE_pp_iid_ip_2}).

Note that the point $n=1001$ is excluded from the data used for LC, LLH and LLM estimation at the boundary point. Similarly the point $n=200$ is excluded
for the case of estimation at the internal point.

From results on these iid regression datasets it can be seen that for boundary value estimation the estimator based on 
$\bar {F}^{LLM}_{x}(y)$ has superior performance as compared to both $\bar {F}_{x}(y)$ and $\bar {F}^{LLH}_{x}(y)$. The
improvement is seen over a wide range of selected bandwidths using both the mean values of the Kolmogorov-Smirnov
test statistic (Tables \ref{DE_pp_iid_bp_1} and \ref{DE_pp_iid_bp_2}) and mean-square error of point prediction
(Tables  \ref{MF_pp_iid_bp_1} and \ref{MF_pp_iid_bp_2}).
Moreover the overall best performance over the selected bandwidth range  from $10,\ldots,140$ is obtained using the Monotone Local Linear Estimator $\bar {F}^{LLM}_{x}(y)$. In addition it can be seen from the plots of the estimated quantiles at  $\alpha = 0.1$ and $\alpha=0.9$ in the boundary case
that the center of the estimated quantile distribution for LLM is aligned more closely to the 'true' quantile value calculated from the simulated data
as shown by the dotted line (Figures \ref{figure_quantile_01_1S} and \ref{figure_quantile_09_1S}).

For the case of estimation at internal points no appreciable differences in performance are noticeable between the 3 estimators
using both the mean values of the Kolmogorov-Smirnov test statistic (Tables \ref{DE_pp_iid_ip_1} and \ref{DE_pp_iid_ip_2}) and 
also using mean-square error of point prediction (Tables \ref{MF_pp_iid_ip_1} and \ref{MF_pp_iid_ip_2}). Similar trends
are noticeable in the quantile plots where the estimated quantiles using LC, LLH and LLM nearly overlap for the internal case
(Figures  \ref{figure_quantile_01_2S} and \ref{figure_quantile_09_2S}).

It can also be seen from Tables \ref{MF_pp_iid_bp_1}, \ref{MF_pp_iid_ip_1},  \ref{MF_pp_iid_bp_2} and \ref{MF_pp_iid_ip_2} that across
the range of bandwidths considered there is negligible loss in best point prediction performance of LLM versus that of LL.

% All of the above studied performance measures have been estimated using 500 realizations of the simulated datasets.

% \footnote{in the density plots you have MF-LC, MF-LL and MF-LLFM;
% I guess MF-LL is LLH (Hansen) and MF-LLFM is LLM the monotone, right?
% Also, the fact that MF-LC is so far off is probably an indication that
% the bandwidth used was not right for it (LC has different optimal bandwidth
% than LL at the boundary); if yes, please state it in the discussion }

% \bigskip

\begin{table}
\centering
\caption{Mean values of KS test statistic over i.i.d. data at boundary point ($n=1001, \tau=0.1$)}
\label{DE_pp_iid_bp_1}
\begin{tabular}{| w | w | w | g |}
\hline
Bandwidth & KS-LC & KS-LLH & KS-LLM \\
\hline
10 & 0.23508 & 0.252884 & 0.275132 \\
\hline
20 & 0.241992 & 0.233996 & 0.23606 \\
\hline
30 & 0.2767 & 0.232064 & 0.218948 \\
\hline
40 & 0.31528 & 0.240476 & 0.20744 \\
\hline
50 & 0.349924 & 0.2554 & 0.2009 \\
\hline
60 & 0.38438 & 0.273648 & 0.204404 \\
\hline
70 & 0.418316 & 0.288032 & 0.21502 \\
\hline
80 & 0.448772 & 0.307672 & 0.231588 \\
\hline
90 & 0.474796 & 0.326224 & 0.253472 \\
\hline
100 & 0.502768 & 0.342884 & 0.275936 \\
\hline
110 & 0.5264 & 0.360888 & 0.2993 \\
\hline
120 & 0.54664 & 0.37786 & 0.320348 \\
\hline
130 & 0.56692 & 0.393392 & 0.34248 \\
\hline
140 & 0.58646 & 0.407108 & 0.359404 \\
\hline
\end{tabular}
\end{table}

\begin{table}
\centering
\caption{Mean values of KS test statistic over i.i.d. data at internal point ($n=200, \tau=0.1$)}
\label{DE_pp_iid_ip_1}
\begin{tabular}{| w | w | w | g |}
\hline
Bandwidth & KS-LC & KS-LLH & KS-LLM \\
\hline
10 & 0.212296 & 0.213792 & 0.213712 \\
\hline
20 & 0.201892 & 0.203264 & 0.203704 \\
\hline
30 & 0.197736 & 0.198904 & 0.197828 \\
\hline
40 & 0.19782 & 0.197296 & 0.196772 \\
\hline
50 & 0.19606 & 0.1949 & 0.19684 \\
\hline
60 & 0.200164 & 0.198304 & 0.198556 \\
\hline
70 & 0.202644 & 0.201472 & 0.202208 \\
\hline
80 & 0.206016 & 0.20534 & 0.207628 \\
\hline
90 & 0.21412 & 0.212608 & 0.21422 \\
\hline
100 & 0.220084 & 0.221096 & 0.2204 \\
\hline
110 & 0.23078 & 0.23064 & 0.231744 \\
\hline
120 & 0.240556 & 0.238724 & 0.240032 \\
\hline
130 & 0.250116 & 0.250692 & 0.250972 \\
\hline
140 & 0.260864 & 0.260696 & 0.259292 \\
\hline
\end{tabular}
\end{table}

\begin{table}
\centering
\caption{Mean values of KS test statistic over i.i.d. data at boundary point ($n=1001, \tau=0.3$)}
\label{DE_pp_iid_bp_2}
\begin{tabular}{| w | w | w | g |}
\hline
Bandwidth & KS-LC & KS-LLH & KS-LLM \\
\hline
10 & 0.207104 & 0.303696 & 0.352912 \\
\hline
20 & 0.148964 & 0.210324 & 0.250856 \\
\hline
30 & 0.125284 & 0.171268 & 0.2058 \\
\hline
40 & 0.112412 & 0.15016 & 0.182176 \\
\hline
50 & 0.107232 & 0.136612 & 0.16702 \\
\hline
60 & 0.107764 & 0.127176 & 0.154944 \\
\hline
70 & 0.111144 & 0.121408 & 0.145624 \\
\hline
80 & 0.119836 & 0.115008 & 0.136968 \\
\hline
90 & 0.126996 & 0.110716 & 0.128792 \\
\hline
100 & 0.137376 & 0.108468 & 0.121452 \\
\hline
110 & 0.14676 & 0.105504 & 0.1165 \\
\hline
120 & 0.157364 & 0.107432 & 0.111452 \\
\hline
130 & 0.165528 & 0.108692 & 0.107532 \\
\hline
140 & 0.175852 & 0.110228 & 0.103772 \\
\hline
\end{tabular}
\end{table}

\begin{table}
\centering
\caption{Mean values of KS test statistic over i.i.d. data at internal point ($n=200, \tau=0.3$)}
\label{DE_pp_iid_ip_2}
\begin{tabular}{| w | w | w | g |}
\hline
Bandwidth & KS-LC & KS-LLH & KS-LLM \\
\hline
10 & 0.152968 & 0.15334 & 0.152252 \\
\hline
20 & 0.119528 & 0.117216 & 0.118916 \\
\hline
30 & 0.103412 & 0.104188 & 0.104388 \\
\hline
40 & 0.097028 & 0.097544 & 0.097348 \\
\hline
50 & 0.0897 & 0.089944 & 0.090576 \\
\hline
60 & 0.0868 & 0.086116 & 0.087244 \\
\hline
70 & 0.083068 & 0.083164 & 0.084304 \\
\hline
80 & 0.082208 & 0.081544 & 0.081452 \\
\hline
90 & 0.080592 & 0.081848 & 0.081572 \\
\hline
100 & 0.07958 & 0.08006 & 0.078328 \\
\hline
110 & 0.080208 & 0.080568 & 0.079604 \\
\hline
120 & 0.08194 & 0.08094 & 0.082332 \\
\hline
130 & 0.082628 & 0.08288 & 0.082256 \\
\hline
140 & 0.084188 & 0.08518 & 0.086076 \\
\hline
\end{tabular}
\end{table}

\begin{table}
\centering
\caption{Point Prediction for Boundary Value over i.i.d. data ($n=1001, \tau=0.1$)}
\label{MF_pp_iid_bp_1}
\scalebox{0.7}{
\begin{tabular}{| w | g | w | g | w | g | w | g | w | g |}
\hline
 Ban & Bias-LC & MSE-LC & Bias-LLH & MSE-LLH & Bias-LLM & MSE-LLM & Bias-LL & MSE-LL \\
 \hline
 10 & -0.01887676 & 0.01265856 & -0.0087034 & 0.01453471 & 0.0004694887 & 0.01667712 & 0.00279478 & 0.01713243 \\
\hline
20 & -0.03782673 & 0.01261435 & -0.01818502 & 0.0126929 & 0.0005444976 & 0.01323652 & 0.003247646 & 0.01340418 \\
\hline
30 & -0.05753609 & 0.01418224 & -0.02725602 & 0.01232877 & -0.001022256 & 0.01200918 & 0.0039133 & 0.01219628 \\
\hline
40 & -0.07724901 & 0.01672728 & -0.03718728 & 0.01259729 & -0.005397138 & 0.01148354 & 0.00354838 & 0.01167496 \\
\hline
50 & -0.09692561 & 0.0200906 & -0.04758345 & 0.01327841 & -0.01222596 & 0.01130622 & 0.002834568 & 0.01139095 \\
\hline
60 & -0.116533 & 0.02423279 & -0.05831195 & 0.01431087 & -0.02106315 & 0.01142789 & 0.002008806 & 0.01120327 \\
\hline
70 & -0.1359991 & 0.02911512 & -0.06918129 & 0.0156254 & -0.03138586 & 0.01185914 & 0.001102312 & 0.01106821 \\
\hline
80 & -0.1555938 & 0.03480583 & -0.08021998 & 0.01722284 & -0.04274234 & 0.01263368 & 8.912064e-05 & 0.01096947 \\
\hline
90 & -0.1752324 & 0.04128715 & -0.09144259 & 0.01910772 & -0.05473059 & 0.01375585 & -0.001070282 & 0.01089842 \\
\hline
100 & -0.1947342 & 0.04848954 & -0.1027918 & 0.02127558 & -0.0670785 & 0.01521865 & -0.002416635 & 0.01084951 \\
\hline
110 & -0.2145001 & 0.05656322 & -0.1142845 & 0.02374615 & -0.07967838 & 0.01704094 & -0.003988081 & 0.01081946 \\
\hline
120 & -0.2343967 & 0.06548142 & -0.1259372 & 0.02651703 & -0.09236019 & 0.01919461 & -0.005818943 & 0.01080699 \\
\hline
130 & -0.2543523 & 0.07522469 & -0.1377167 & 0.02960364 & -0.1050934 & 0.02168698 & -0.007939144 & 0.01081259 \\
\hline
140 & -0.2740635 & 0.08563245 & -0.1496325 & 0.03301117 & -0.1178388 & 0.02451228 & -0.01037417 & 0.01083832 \\
\hline
\end{tabular}}
\end{table}

\begin{table}
\centering
\caption{Point Prediction for Internal Value over i.i.d. data ($n=200, \tau=0.1$)}
\label{MF_pp_iid_ip_1}
\scalebox{0.7}{
\begin{tabular}{| w | g | w | g | w | g | w | g | w | g | }
\hline
 Ban & Bias-LC & MSE-LC & Bias-LLH & MSE-LLH & Bias-LLM & MSE-LLM & Bias-LL & MSE-LL \\
 \hline
10 & 0.005693694 & 0.01026982 & 0.005815108 & 0.01027252 & 0.005811741 & 0.01027231 & 0.005672309 & 0.01027341 \\
\hline
20 & 0.004548762 & 0.009868812 & 0.004644668 & 0.009871222 & 0.004640743 & 0.009871005 & 0.004547984 & 0.009883257 \\
\hline
30 & 0.003077572 & 0.009736622 & 0.003193559 & 0.009739295 & 0.003189924 & 0.009738919 & 0.003108078 & 0.009754927 \\
\hline
40 & 0.001168265 & 0.009684642 & 0.001329604 & 0.009685997 & 0.001325573 & 0.009685696 & 0.001205735 & 0.009703492 \\
\hline
50 & -0.001163283 & 0.009671566 & -0.0009392514 & 0.009670138 & -0.0009440976 & 0.009670008 & -0.001162398 & 0.009689214 \\
\hline
60 & -0.003874557 & 0.009682447 & -0.00359328 & 0.009680945 & -0.003598744 & 0.009680969 & -0.003997042 & 0.009703 \\
\hline
70 & -0.006944759 & 0.009723612 & -0.006615935 & 0.009717111 & -0.006621406 & 0.009717225 & -0.007307346 & 0.009745675 \\
\hline
80 & -0.01035534 & 0.009789969 & -0.009987875 & 0.009781065 & -0.009992804 & 0.009781194 & -0.01109961 & 0.009822695 \\
\hline
90 & -0.01407319 & 0.009888265 & -0.01368629 & 0.009877023 & -0.01369037 & 0.009877157 & -0.01537421 & 0.009942768 \\
\hline
100 & -0.01808254 & 0.01002258 & -0.01768867 & 0.01001026 & -0.01769184 & 0.01001041 & -0.02012788 & 0.01011708 \\
\hline
110 & -0.02234318 & 0.01020278 & -0.02197526 & 0.01018668 & -0.02197765 & 0.01018686 & -0.02535515 & 0.01035866 \\
\hline
120 & -0.02686568 & 0.01042781 & -0.02652964 & 0.01041258 & -0.02653147 & 0.0104128 & -0.03104801 & 0.0106819 \\
\hline
130 & -0.03163397 & 0.01071166 & -0.03133849 & 0.01069454 & -0.03133999 & 0.01069479 & -0.03719388 & 0.01110199 \\
\hline
140 & -0.03662567 & 0.01105637 & -0.03639079 & 0.01103926 & -0.03639212 & 0.01103955 & -0.04377252 & 0.0116341 \\
\hline
\end{tabular}}
\end{table}

\begin{table}
\centering
\caption{Point Prediction for Boundary Value over i.i.d. data ($n=1001, \tau=0.3$)}
\label{MF_pp_iid_bp_2}
\scalebox{0.7}{
\begin{tabular}{| w | g | w | g | w | g | w | g | w | g | }
\hline
Ban & Bias-LC & MSE-LC & Bias-LLH & MSE-LLH & Bias-LLM & MSE-LLM & Bias-LL & MSE-LL \\
\hline
10 & 0.04888178 & 0.301925 & 0.07073083 & 0.3540897 & 0.07920865 & 0.384878 & 0.0808868 & 0.4035579 \\
\hline
20 & 0.02525561 & 0.2802656 & 0.05074344 & 0.3037839 & 0.06735271 & 0.3233827 & 0.07335949 & 0.3276234 \\
\hline
30 & 0.00374298 & 0.2731737 & 0.038811 & 0.2892723 & 0.06222332 & 0.3013529 & 0.07053577 & 0.3027942 \\
\hline
40 & -0.01695169 & 0.270537 & 0.02715055 & 0.2805281 & 0.05849931 & 0.2922475 & 0.06822172 & 0.293552 \\
\hline
50 & -0.03718522 & 0.2696291 & 0.01614152 & 0.2761872 & 0.05612087 & 0.2867147 & 0.06656515 & 0.2892179 \\
\hline
60 & -0.05753523 & 0.2699832 & 0.005048478 & 0.2739688 & 0.05384922 & 0.2829767 & 0.0649322 & 0.2860192 \\
\hline
70 & -0.07760465 & 0.271603 & -0.005574987 & 0.2723361 & 0.0513094 & 0.2798923 & 0.06320544 & 0.2830793 \\
\hline
80 & -0.09765073 & 0.2742877 & -0.01633413 & 0.271128 & 0.04834131 & 0.2770397 & 0.06143642 & 0.2803242 \\
\hline
90 & -0.1176859 & 0.2780296 & -0.02722356 & 0.2704552 & 0.04514186 & 0.2748099 & 0.05960562 & 0.2778554 \\
\hline
100 & -0.1373472 & 0.2827116 & -0.0383895 & 0.2701542 & 0.04137961 & 0.2727937 & 0.05763437 & 0.2757286 \\
\hline
110 & -0.1572939 & 0.2883236 & -0.04971082 & 0.2703248 & 0.03701344 & 0.2709994 & 0.05544761 & 0.2739321 \\
\hline
120 & -0.1769863 & 0.294608 & -0.0611495 & 0.2709176 & 0.03212707 & 0.2695289 & 0.0530012 & 0.2724221 \\
\hline
130 & -0.1965911 & 0.3018083 & -0.07255455 & 0.2717088 & 0.02680826 & 0.2683285 & 0.05027668 & 0.2711495 \\
\hline
140 & -0.2158054 & 0.3097015 & -0.08401642 & 0.2728317 & 0.02098977 & 0.2673724 & 0.04726651 & 0.2700701 \\
\hline
 \end{tabular}}
\end{table}

\begin{table}
\centering
\caption{Point Prediction for Internal Value over i.i.d. data ($n=200, \tau=0.3$)}
\label{MF_pp_iid_ip_2}
\scalebox{0.7}{
\begin{tabular}{| w | g | w | g | w | g | w | g | w | g | }
\hline
Ban & Bias-LC & MSE-LC & Bias-LLH & MSE-LLH & Bias-LLM & MSE-LLM & Bias-LL & MSE-LL \\
\hline
10 & 0.009184716 & 0.2520511 & 0.01220923 & 0.2521932 & 0.01220434 & 0.2521952 & 0.007409901 & 0.2516582 \\
\hline
20 & 0.01372525 & 0.2431836 & 0.01526585 & 0.2435718 & 0.01526117 & 0.2435718 & 0.01263826 & 0.2426903 \\
\hline
30 & 0.0148307 & 0.2398743 & 0.01582708 & 0.2401292 & 0.01582349 & 0.2401341 & 0.01395436 & 0.2395701 \\
\hline
40 & 0.0135934 & 0.2381523 & 0.01432564 & 0.2382689 & 0.01432314 & 0.2382728 & 0.01288775 & 0.2379284 \\
\hline
50 & 0.01125721 & 0.236852 & 0.011912 & 0.2369737 & 0.01190766 & 0.2369759 & 0.01078182 & 0.2367428 \\
\hline
60 & 0.008293956 & 0.2359636 & 0.008883749 & 0.2359976 & 0.008879099 & 0.2360007 & 0.007971824 & 0.2358225 \\
\hline
70 & 0.004809638 & 0.2352631 & 0.005346559 & 0.2352719 & 0.005342764 & 0.235277 & 0.004580992 & 0.235121 \\
\hline
80 & 0.0009735356 & 0.2347759 & 0.001361408 & 0.2347516 & 0.001357999 & 0.2347585 & 0.0006901448 & 0.2346118 \\
\hline
90 & -0.003467449 & 0.234453 & -0.003042608 & 0.2344041 & -0.003046705 & 0.2344117 & -0.00365963 & 0.2342717 \\
\hline
100 & -0.008232451 & 0.2342181 & -0.007859816 & 0.2342051 & -0.007864671 & 0.2342125 & -0.008456445 & 0.2340811 \\
\hline
110 & -0.01347908 & 0.2341583 & -0.01309377 & 0.2341384 & -0.01309954 & 0.234145 & -0.01370081 & 0.2340256 \\
\hline
120 & -0.01912791 & 0.2342317 & -0.01874779 & 0.2341951 & -0.01875384 & 0.2342009 & -0.01939379 & 0.2340969 \\
\hline
130 & -0.02516629 & 0.2344631 & -0.0248178 & 0.2343727 & -0.02482374 & 0.234378 & -0.02553028 & 0.2342927 \\
\hline
140 & -0.0316367 & 0.2347946 & -0.0312908 & 0.2346738 & -0.03129606 & 0.2346788 & -0.03209508 & 0.2346152 \\
\hline
\end{tabular}}
\end{table}

 \graphicspath{{/Users/rumpagiri/Documents/NONPARAMETRIC/model_free/papers/REGRESSION/v2_011817}}
\DeclareGraphicsExtensions{.png}

{\begin{figure}[!t]
  \centering
  \includegraphics[width=4.5in, height=3.5in]{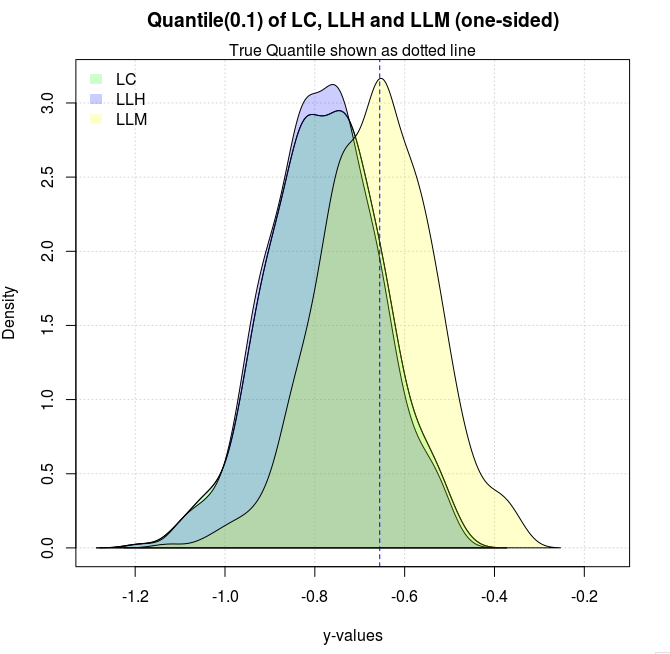}
  \caption{Estimated versus true quantile values ($\alpha=0.1$) for 1-sided estimation, i.i.d. errors ($\tau = 0.3$)}
  \label{figure_quantile_01_1S}
\end{figure}}

\graphicspath{{/Users/rumpagiri/Documents/NONPARAMETRIC/model_free/papers/REGRESSION/v2_011817}}
\DeclareGraphicsExtensions{.png}

{\begin{figure}[!t]
  \centering
  \includegraphics[width=4.5in, height=3.5in]{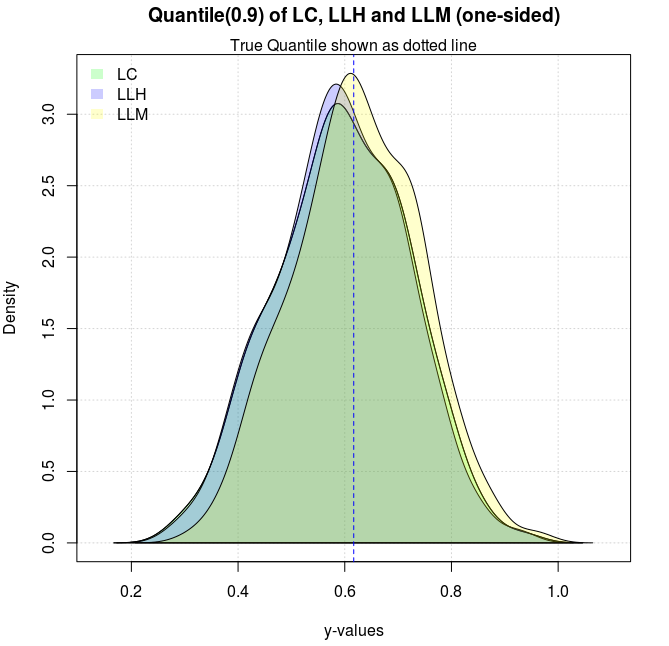}
  \caption{Estimated versus true quantile values ($\alpha=0.9$) for 1-sided estimation, i.i.d. errors ($\tau = 0.3$)}
  \label{figure_quantile_09_1S}
\end{figure}}

\clearpage

\graphicspath{{/Users/rumpagiri/Documents/NONPARAMETRIC/model_free/papers/REGRESSION/v2_011817}}
\DeclareGraphicsExtensions{.png}

{\begin{figure}[!t]
  \centering
  \includegraphics[width=4.5in, height=3.5in]{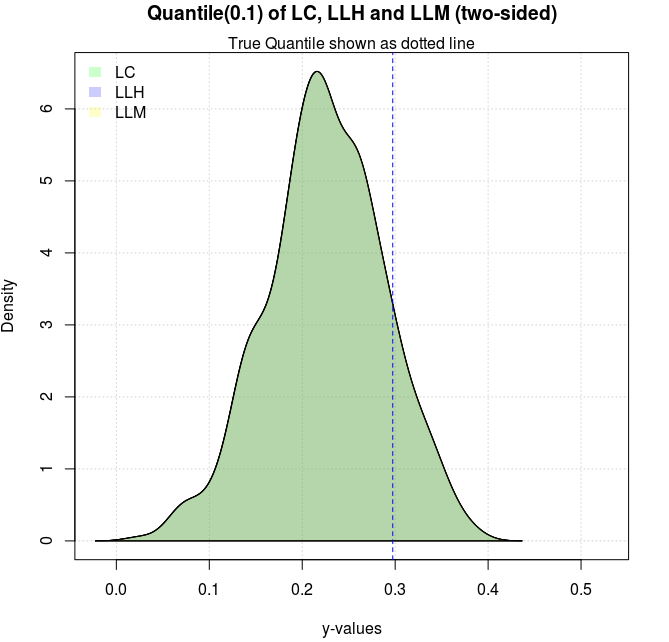}
  \caption{Estimated versus true quantile values ($\alpha=0.1$) for 2-sided estimation, i.i.d. errors ($\tau = 0.3$)}
  \label{figure_quantile_01_2S}
\end{figure}}

\graphicspath{{/Users/rumpagiri/Documents/NONPARAMETRIC/model_free/papers/REGRESSION/v2_011817}}
\DeclareGraphicsExtensions{.png}

{\begin{figure}[!t]
  \centering
  \includegraphics[width=4.5in, height=3.5in]{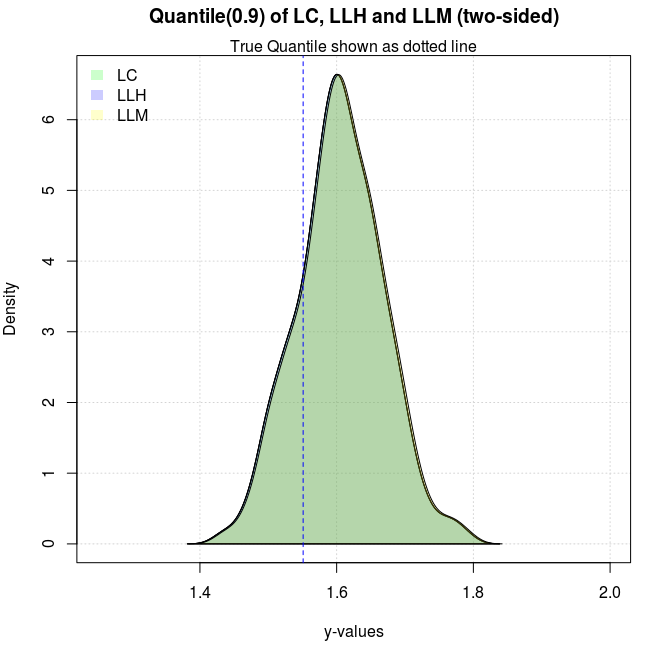}
  \caption{Estimated versus true quantile values ($\alpha=0.9$) for 2-sided estimation, i.i.d. errors ($\tau = 0.3$)}
  \label{figure_quantile_09_2S}
\end{figure}}

%\footnote{There should be a discussion of findings within each subsection--in particular, you need
%to add a small comment regarding each Table and Figure you include. [Of course if finding of Table 
%X is similar to that of Table Y you just say that--but you need to say it]
%
%Also, the last section that
%you called ``Discussion'' is renamed ``Conclusions'' and should be a re-cap of major points,
%  methods and simulation}

\subsection{Simulation: Additive model with heteroskedastic errors}

Data $Y_i$ for $i =1, \ldots, 1001$ were simulated as per model (\ref{MB_eq.model})
with $\mu (x_i)= \sin(x_i)$, $\sigma(x_i)=\tau x_i$ where $x_i = \frac {i} {n}$ 
and the errors $\epsilon_i$ as ~i.i.d. $\frac {1}{2} \chi_2^2 - 1$.
Sample size $n$ was set to $1001$. A total of 500 such realizations were generated for
this study. \\
\indent Results for the mean-value of the Kolmogorov-Smirnov test statistic between the LC, LLH and LLM estimated distributions and empirical distribution calculated using available
values of the simulated data are given in Tables \ref{DE_pp_hetero_bp_1}, \ref{DE_pp_hetero_ip_1}, \ref{DE_pp_hetero_bp_2} and \ref{DE_pp_hetero_ip_2} for boundary point $n=1001$ and internal point $n=200$ for values of $\tau=0.1$ and $0.3$ over a range  of bandwidths: $10, 20, \ldots, 140$.

\indent Point prediction performance values are provided for the same cases in Tables \ref{MF_pp_hetero_bp_1}, \ref{MF_pp_hetero_ip_1}, \ref{MF_pp_hetero_bp_2} and \ref{MF_pp_hetero_ip_2}.

Note that the point $n=1001$ is excluded from the data used for LC, LLH and LLM estimation at the boundary point. Similarly the point $n=200$ is excluded
for the case of estimation at the internal point.

From results on these heteroskedastic regression datasets it can be seen that for boundary value estimation the estimator based on 
$\bar {F}^{LLM}_{x}(y)$ has superior performance as compared to both $\bar {F}_{x}(y)$ and $\bar {F}^{LLH}_{x}(y)$. The
improvement is seen over a wide range of selected bandwidths using both the mean values of the Kolmogorov-Smirnov
test statistic (Tables \ref{DE_pp_hetero_bp_1} and \ref{DE_pp_hetero_bp_2}) and mean-square error of point prediction
(Tables  \ref{MF_pp_hetero_bp_1} and \ref{MF_pp_hetero_bp_2}).
Moreover the overall best performance over the selected bandwidth range  from $10,\ldots,140$ is obtained using the Monotone Local Linear Estimator $\bar {F}^{LLM}_{x}(y)$.

For the case of estimation at internal points no appreciable differences in performance are noticeable between the 3 estimators
using both the mean values of the Kolmogorov-Smirnov test statistic (Tables \ref{DE_pp_hetero_ip_1} and \ref{DE_pp_hetero_ip_2}) and 
also using mean-square error of point prediction (Tables \ref{MF_pp_hetero_ip_1} and \ref{MF_pp_hetero_ip_2}). 

It can also be seen from Tables \ref{MF_pp_hetero_bp_1}, \ref{MF_pp_hetero_ip_1},  \ref{MF_pp_hetero_bp_2} and \ref{MF_pp_hetero_ip_2} that---across
the range of bandwidths considered---there is negligible loss in best point prediction performance of LLM versus that of LL. This finding  is  unexpected   since
it has been widely believed that the LL method   gives optimal point estimators and/or predictors. It appears that the monotonicity correction does not hurt the resulting point  estimators/predictors
which is encouraging.

% All of the above studied performance measures have been estimated using 500 realizations of the simulated datasets.

% \footnote{need a description of the entries of the tables, and then discussion of the
% result}

%\footnote{
% for Tables 5 and later: (i) what is LL-MB method?
%(ii) maybe drop the distinction between P and F for residuals; just keep the
%regular (F) since for point estimation it doesn;t make much difference.
% Also add a remark that for the numerical work the Nadarya-Watson  estimators were computed using the
% alternative formula given in the book. 
%(iii) but for the LL methods I didn;t give an alternative formula
%using residuals... so I am not sure why you have two entries (P and F) here... 
%}
 
% \bigskip

\clearpage

% \FloatBarrier

\begin{table}
\centering
\caption{Mean values of KS test statistic over heteroskedastic data at boundary point ($n=1001, \tau=0.1$)}
\label{DE_pp_hetero_bp_1}
\begin{tabular}{| w | w | w | g |}
\hline
Bandwidth & KS-LC & KS-LLH & KS-LLM \\
\hline
10 & 0.361228 & 0.3619 & 0.368288 \\
\hline
20 & 0.39358 & 0.3606 & 0.336436 \\
\hline
30 & 0.43216 & 0.371372 & 0.326076 \\
\hline
40 & 0.470316 & 0.388952 & 0.325116 \\
\hline
50 & 0.506436 & 0.408316 & 0.335152 \\
\hline
60 & 0.53998 & 0.42548 & 0.350864 \\
\hline
70 & 0.572256 & 0.44356 & 0.371324 \\
\hline
80 & 0.599836 & 0.462808 & 0.393896 \\
\hline
90 & 0.6269 & 0.47816 & 0.415468 \\
\hline
100 & 0.651132 & 0.499376 & 0.44184 \\
\hline
110 & 0.670604 & 0.516304 & 0.462756 \\
\hline
120 & 0.69 & 0.529796 & 0.485004 \\
\hline
130 & 0.706968 & 0.545344 & 0.505352 \\
\hline
140 & 0.72394 & 0.562432 & 0.5257 \\
\hline
\end{tabular}
\end{table}

\begin{table}
\centering
\caption{Mean values of KS test statistic over heteroskedastic data at internal point ($n=200, \tau=0.1$)}
\label{DE_pp_hetero_ip_1}
\begin{tabular}{| w | w | w | g |}
\hline
Bandwidth & KS-LC & KS-LLH & KS-LLM \\
\hline
10 & 0.459776 & 0.461528 & 0.461176 \\
\hline
20 & 0.461872 & 0.462716 & 0.4603 \\
\hline
30 & 0.46576 & 0.467308 & 0.464956 \\
\hline
40 & 0.468904 & 0.471824 & 0.470172 \\
\hline
50 & 0.47436 & 0.475916 & 0.474864 \\
\hline
60 & 0.482716 & 0.482476 & 0.47912 \\
\hline
70 & 0.488952 & 0.488444 & 0.486656 \\
\hline
80 & 0.495916 & 0.495736 & 0.495056 \\
\hline
90 & 0.503672 & 0.503052 & 0.502708 \\
\hline
100 & 0.5105 & 0.513116 & 0.51026 \\
\hline
110 & 0.519052 & 0.518104 & 0.518928 \\
\hline
120 & 0.528456 & 0.528444 & 0.527104 \\
\hline
130 & 0.537336 & 0.536916 & 0.535632 \\
\hline
140 & 0.545264 & 0.545496 & 0.543776 \\
\hline
\end{tabular}
\end{table}

\begin{table}
\centering
\caption{Mean values of KS test statistic over heteroskedastic data at boundary point ($n=1001, \tau=0.3$)}
\label{DE_pp_hetero_bp_2}
\begin{tabular}{| w | w | w | g |}
\hline
Bandwidth & KS-LC & KS-LLH & KS-LLM \\
\hline
10 & 0.208708 & 0.28022 & 0.323664 \\
\hline
20 & 0.176304 & 0.210876 & 0.241228 \\
\hline
30 & 0.178416 & 0.189656 & 0.206996 \\
\hline
40 & 0.189136 & 0.17842 & 0.186628 \\
\hline
50 & 0.204484 & 0.175508 & 0.173096 \\
\hline
60 & 0.220652 & 0.177144 & 0.163916 \\
\hline
70 & 0.240692 & 0.181092 & 0.158476 \\
\hline
80 & 0.25784 & 0.186648 & 0.15736 \\
\hline
90 & 0.277888 & 0.191396 & 0.156008 \\
\hline
100 & 0.295264 & 0.20092 & 0.159028 \\
\hline
110 & 0.312968 & 0.20922 & 0.163296 \\
\hline
120 & 0.330008 & 0.216464 & 0.167872 \\
\hline
130 & 0.345432 & 0.22344 & 0.17522 \\
\hline
140 & 0.36082 & 0.234392 & 0.181376 \\
\hline
\end{tabular}
\end{table}

\begin{table}
\centering
\caption{Mean values of KS test statistic over heteroskedastic data at internal point ($n=200, \tau=0.3$)}
\label{DE_pp_hetero_ip_2}
\begin{tabular}{| w | w | w | g |}
\hline
Bandwidth & KS-LC & KS-LLH & KS-LLM \\
\hline
10 & 0.3289 & 0.329088 & 0.329112 \\
\hline
20 & 0.327172 & 0.326072 & 0.3268 \\
\hline
30 & 0.327236 & 0.32788 & 0.3275 \\
\hline
40 & 0.331784 & 0.3309 & 0.33186 \\
\hline
50 & 0.337856 & 0.337888 & 0.337692 \\
\hline
60 & 0.343504 & 0.344328 & 0.343368 \\
\hline
70 & 0.350048 & 0.351444 & 0.349592 \\
\hline
80 & 0.3588 & 0.359188 & 0.358944 \\
\hline
90 & 0.36826 & 0.368708 & 0.368008 \\
\hline
100 & 0.378308 & 0.376472 & 0.377692 \\
\hline
110 & 0.386636 & 0.3864 & 0.388256 \\
\hline
120 & 0.39642 & 0.395744 & 0.39754 \\
\hline
130 & 0.4055 & 0.408072 & 0.40714 \\
\hline
140 & 0.418516 & 0.4171 & 0.41794 \\
\hline
\end{tabular}
\end{table}

\begin{table}
\centering
\caption{Point Prediction for Boundary Value over heteroskedastic data ($n=1001, \tau=0.1$)}
\label{MF_pp_hetero_bp_1}
\scalebox{0.7}{
\begin{tabular}{| w | g | w | g | w | g | w | g | w | g |}
\hline
Ban & Bias-LC & MSE-LC & Bias-LLH & MSE-LLH & Bias-LLM & MSE-LLM & Bias-LL & MSE-LL \\
\hline
10 & -0.01646515 & 0.0110308 & -0.008415339 & 0.01301928 & 0.003362834 & 0.01521503 & 0.002122231 & 0.01532911 \\
\hline
20 & -0.03418985 & 0.01113183 & -0.01803682 & 0.01111592 & 0.001465109 & 0.01164382 & 0.003045892 & 0.01196587 \\
\hline
30 & -0.05291763 & 0.01251871 & -0.02791687 & 0.0110065 & -0.001493594 & 0.01066538 & 0.003162759 & 0.01102039 \\
\hline
40 & -0.07217657 & 0.01484132 & -0.03844108 & 0.01144334 & -0.007355843 & 0.01025364 & 0.003252626 & 0.01051494 \\
\hline
50 & -0.09186859 & 0.0180368 & -0.0493472 & 0.01222871 & -0.01604004 & 0.01020275 & 0.003291589 & 0.01020678 \\
\hline
60 & -0.1116673 & 0.02205503 & -0.06052473 & 0.01337097 & -0.0266163 & 0.0105145 & 0.003183081 & 0.01002049 \\
\hline
70 & -0.1312554 & 0.02681084 & -0.07204081 & 0.01484635 & -0.03845618 & 0.01120131 & 0.002843088 & 0.009915576 \\
\hline
80 & -0.1512692 & 0.03246252 & -0.08373385 & 0.01662921 & -0.05099656 & 0.01226805 & 0.002239256 & 0.009858149 \\
\hline
90 & -0.1714417 & 0.03896746 & -0.09557852 & 0.01872622 & -0.06394962 & 0.01372077 & 0.00136753 & 0.009824624 \\
\hline
100 & -0.1916003 & 0.04627765 & -0.1075785 & 0.02114855 & -0.07708492 & 0.01554708 & 0.0002256174 & 0.009802568 \\
\hline
110 & -0.2119687 & 0.05448537 & -0.1197012 & 0.02389638 & -0.09028337 & 0.01774215 & -0.001196002 & 0.009787441 \\
\hline
120 & -0.2326798 & 0.06368262 & -0.1320067 & 0.02699023 & -0.1035047 & 0.0202921 & -0.002912961 & 0.009779257 \\
\hline
130 & -0.2535364 & 0.07381161 & -0.1444581 & 0.03043434 & -0.1167033 & 0.02319127 & -0.004943721 & 0.009780505 \\
\hline
140 & -0.2740579 & 0.08462823 & -0.1570383 & 0.03422973 & -0.1299138 & 0.02644559 & -0.007307095 & 0.009795173 \\
\hline
\end{tabular}}
\end{table}

\begin{table}
\centering
\caption{Point Prediction for Internal Value over heteroskedastic data ($n=200, \tau=0.1$)}
\label{MF_pp_hetero_ip_1}
\scalebox{0.7}{
\begin{tabular}{| w | g | w | g | w | g | w | g | w | g |}
\hline
Ban & Bias-LC & MSE-LC & Bias-LLH & MSE-LLH & Bias-LLM & MSE-LLM & Bias-LL & MSE-LL \\
\hline
10 & -0.00078446 & 0.0004397085 & -0.001282314 & 0.0004403816 & -0.001281847 & 0.0004403461 & -0.001460641 & 0.0004417506 \\
\hline
20 & -0.001122367 & 0.0004306633 & -0.001431476 & 0.0004311207 & -0.001431238 & 0.0004311334 & -0.001977922 & 0.0004335427 \\
\hline
30 & -0.002288569 & 0.0004309951 & -0.002426097 & 0.0004313394 & -0.002424798 & 0.0004313337 & -0.003182182 & 0.0004360195 \\
\hline
40 & -0.00405804 & 0.0004390668 & -0.004017686 & 0.0004388123 & -0.004015654 & 0.0004387818 & -0.004960882 & 0.0004476175 \\
\hline
50 & -0.006300199 & 0.0004597561 & -0.006090049 & 0.0004573097 & -0.006086971 & 0.0004572689 & -0.007269184 & 0.0004732545 \\
\hline
60 & -0.008952297 & 0.0004986956 & -0.00857106 & 0.0004917471 & -0.008566653 & 0.0004916759 & -0.01008903 & 0.0005201175 \\
\hline
70 & -0.01195461 & 0.0005599192 & -0.0114063 & 0.0005469568 & -0.01140055 & 0.0005468245 & -0.01341156 & 0.0005966537 \\
\hline
80 & -0.01524307 & 0.000648231 & -0.01455151 & 0.0006275842 & -0.01454456 & 0.0006273696 & -0.01723004 & 0.000712577 \\
\hline
90 & -0.0188042 & 0.0007686332 & -0.0179713 & 0.0007381116 & -0.01796344 & 0.0007378019 & -0.02153766 & 0.0008788525 \\
\hline
100 & -0.02260511 & 0.0009254938 & -0.02163909 & 0.0008829351 & -0.02163065 & 0.000882528 & -0.02632699 & 0.00110763 \\
\hline
110 & -0.02662906 & 0.001123084 & -0.02553604 & 0.001066478 & -0.02552742 & 0.001065985 & -0.03158974 & 0.001412141 \\
\hline
120 & -0.03085926 & 0.001365925 & -0.02964955 & 0.001293297 & -0.02964117 & 0.00129274 & -0.03731584 & 0.001806523 \\
\hline
130 & -0.03531386 & 0.001660546 & -0.03397167 & 0.001568158 & -0.03396393 & 0.00156757 & -0.0434914 & 0.002305438 \\
\hline
140 & -0.03995794 & 0.002010171 & -0.0384976 & 0.001896071 & -0.03849081 & 0.00189549 & -0.05009551 & 0.002923419 \\
\hline
\end{tabular}}
\end{table}

\begin{table}
\centering
\caption{Point Prediction for Boundary Value over heteroskedastic data ($n=1001, \tau=0.3$)}
\label{MF_pp_hetero_bp_2}
\scalebox{0.7}{
\begin{tabular}{| w | g | w | g | w | g | w | g | w | g |}
\hline
Ban & Bias-LC & MSE-LC & Bias-LLH & MSE-LLH & Bias-LLM & MSE-LLM & Bias-LL & MSE-LL \\
\hline
0 & -0.01641585 & 0.273216 & -0.01371259 & 0.3278422 & 0.01500573 & 0.3662851 & 0.01063269 & 0.3832281 \\
\hline
20 & -0.02085331 & 0.2520507 & -0.0253276 & 0.274159 & 0.002731055 & 0.2896534 & 0.01538866 & 0.2991516 \\
\hline
30 & -0.02981426 & 0.2462187 & -0.03060796 & 0.2589025 & 0.003270715 & 0.2685365 & 0.0163369 & 0.2755266 \\
\hline
40 & -0.04068759 & 0.2442488 & -0.03742699 & 0.2526514 & 0.002433103 & 0.2586642 & 0.01748551 & 0.2629147 \\
\hline
50 & -0.05443176 & 0.2442541 & -0.04586018 & 0.2488821 & 0.0005299281 & 0.2526573 & 0.01882287 & 0.2552529 \\
\hline
60 & -0.06977487 & 0.245767 & -0.05475483 & 0.2474683 & -0.001728694 & 0.248843 & 0.0199724 & 0.2506579 \\
\hline
70 & -0.08589639 & 0.2481108 & -0.06470145 & 0.2471712 & -0.005360827 & 0.2463975 & 0.02061821 & 0.2481124 \\
\hline
80 & -0.1036357 & 0.25121 & -0.07550857 & 0.2474051 & -0.01066184 & 0.2448518 & 0.02070028 & 0.2467569 \\
\hline
90 & -0.1221155 & 0.2551902 & -0.08684923 & 0.2482818 & -0.01739231 & 0.2440367 & 0.02029725 & 0.2459808 \\
\hline
100 & -0.1410488 & 0.2599296 & -0.09877418 & 0.2499431 & -0.02522804 & 0.2438554 & 0.01949336 & 0.2454429 \\
\hline
110 & -0.1599352 & 0.2653016 & -0.1111362 & 0.252298 & -0.03400529 & 0.2440469 & 0.0183332 & 0.2449864 \\
\hline
120 & -0.1798873 & 0.2718873 & -0.1241105 & 0.2552008 & -0.04372748 & 0.2444396 & 0.01682687 & 0.2445524 \\
\hline
130 & -0.2001088 & 0.2793124 & -0.1376482 & 0.2586551 & -0.05435831 & 0.2450597 & 0.01496614 & 0.2441256 \\
\hline
140 & -0.2196351 & 0.2872558 & -0.1514669 & 0.2625969 & -0.06555938 & 0.2460242 & 0.01273652 & 0.2437067 \\
\hline
\end{tabular}}
\end{table}

\begin{table}
\centering
\caption{Point Prediction for Internal Value over heteroskedastic data ($n=200, \tau=0.3$)}
\label{MF_pp_hetero_ip_2}
\scalebox{0.7}{
\begin{tabular}{| w | g | w | g | w | g | w | g | w | g |}
\hline
Ban & Bias-LC & MSE-LC & Bias-LLH & MSE-LLH & Bias-LLM & MSE-LLM & Bias-LL & MSE-LL \\
\hline
10 & -0.005989017 & 0.01091506 & -0.009105295 & 0.01090718 & -0.009100397 & 0.01090687 & -0.006151798 & 0.01102828 \\
\hline
20 & -0.004317512 & 0.01067232 & -0.006852094 & 0.01066366 & -0.006845515 & 0.01066378 & -0.005549238 & 0.01077156 \\
\hline
30 & -0.004591794 & 0.01059617 & -0.006678386 & 0.0105944 & -0.006665835 & 0.01059435 & -0.006333703 & 0.01068745 \\
\hline
40 & -0.005937551 & 0.01054613 & -0.007674744 & 0.01055486 & -0.007656429 & 0.01055456 & -0.00795147 & 0.01063841 \\
\hline
50 & -0.007974463 & 0.01051967 & -0.009442124 & 0.01053534 & -0.009416436 & 0.01053501 & -0.01019012 & 0.01061418 \\
\hline
60 & -0.01058953 & 0.01053145 & -0.01180495 & 0.01054554 & -0.01176999 & 0.01054509 & -0.01297439 & 0.01062656 \\
\hline
70 & -0.01373489 & 0.01057533 & -0.01467215 & 0.01059193 & -0.01462675 & 0.01059104 & -0.01627809 & 0.01068457 \\
\hline
80 & -0.01725266 & 0.01066128 & -0.01798338 & 0.01067947 & -0.01792693 & 0.01067781 & -0.02008891 & 0.01079614 \\
\hline
90 & -0.02118215 & 0.0107964 & -0.02169107 & 0.01081295 & -0.02162338 & 0.01081019 & -0.0243973 & 0.01096978 \\
\hline
100 & -0.02546816 & 0.01098609 & -0.02575577 & 0.01099723 & -0.02567729 & 0.01099311 & -0.0291937 & 0.01121525 \\
\hline
110 & -0.03007643 & 0.01123397 & -0.0301445 & 0.01123745 & -0.03005627 & 0.01123178 & -0.03446792 & 0.01154379 \\
\hline
120 & -0.03496193 & 0.01154587 & -0.03483024 & 0.01153901 & -0.03473374 & 0.01153169 & -0.04020819 & 0.01196797 \\
\hline
130 & -0.04015664 & 0.01193249 & -0.03979071 & 0.01190758 & -0.03968792 & 0.01189862 & -0.04639914 & 0.0125013 \\
\hline
140 & -0.04561132 & 0.01240015 & -0.04500812 & 0.01234911 & -0.04490124 & 0.01233864 & -0.05301874 & 0.01315745 \\
\hline
\end{tabular}}
\end{table}

\subsection{Real-life example: Wage dataset}

The {\tt Wage} dataset from the {\tt ISLR} package \cite{James2013} was selected as a real-life example to demonstrate the differences in estimated local densities
estimated using the LC, LLH and LLM methods.
% local constant, local linear (as per Hansen's proposal) and monotone local linear density estimation.
% \footnote{rename LLH (Hansen) and   LLM }
The full dataset has 3000 points and has been constructed from the Current Population Survey (CPS) data for year 2011.  Point Prediction is used as the criterion
for demonstrating performance differences between the three distribution estimators. This dataset is an example
of regression data distributed non-uniformly and hence the local linear estimator (LL) based on equations \ref{NSTS.eq.locallinearF} and \ref{NSTS.eq.locallinearweights}
is expected to give the best performance in such cases. However our study involves using point-prediction
using the three distribution estimators $\bar {F}_{x}(y)$, $\bar {F}^{LLH}_{x}(y)$ or $\bar {F}^{LLM}_{x}(y)$.
Among these 3 estimators LLM gives the best point prediction performance and we show that using 
this estimator causes negligible loss in performance compared to using LL.

% \footnote{you need to state that for point prediction LL is best and LLM/LLH are not needed;
% what we are aiming to show is that there is not much loss in efficiency by using LLM even here}

From the plot of the dataset in Figure \ref{ISLR_data} with superimposed 
smoother (obtained using {\tt loess} fitting from the R package {\tt lattice})
% \footnote{say how you got the smoother}
 it can be noted
that the regression function is sloping upwards at the left boundary whereas it flattens out
at the right boundary. Hence, at the right boundary, local constant methods suffice and
 should be practically equivalent to local linear methods. 
The left boundary is more interesting, and this is where our 
numerical work will focus. 
To carry this out, we created a second version of the data where logwage
 is tabulated versus decreasing age and performed point prediction over the last 231 values of this
backward dataset, i.e., the first 231 values of the original. 
Since this is a regression dataset with non-uniformly distributed design points we determine bandwidths for LC, LLH and LLM  using the
2-sided predictive cross-validation procedure outlined in Section \ref{MF_bw}. We predict the value of logwage at $i$ and 
compare it with the known value at that point where $i=2770,\ldots,3000$ to determine the MSE of point prediction.
% on  boundary points (2700:3000) based on the previous observations using all 3 model-free methods does not show much diff
% is shown in Table \ref{ISLR_Wage_bp}.
Plots of the conditional density function estimated using the three model-free methods LC, LLH and LLM at a selected point are  shown in Figure \ref{de_islr_density_figure}
along with that of LL as reference. 

Point prediction results for all three methods over data points $2770,\ldots,3000$ (logwage versus decreasing  age) are given in Table \ref{ISLR_Wage_bp}. It can be seen from this table that LLM has the best point prediction performance and this closely matches that of LL. As in the case of simulated data,
 this is an unexpected and encouraging result indicating that the LLM distribution may be an all-around
favorable estimator both in terms of its quantiles as well as its center of location used for point estimation and  prediction purposes. 

% The close matching
%is also confirmed by Figure \ref{de_islr_density_figure} where the curves for LLM and LL are nearly %overlapping.
 
% \footnote{you need to include regular LL both    in Figure \ref{de_islr_density_figure} and in Table \ref{ISLR_Wage_bp}.} 

%It can be seen that more structure
%is present in the plots for the local linear methods as compared to the estimation using
%the local constant approach.
%\footnote{more structure can just mean undersmoothing--not nec. good}
%\footnote{the three density figures look  identical side by side; can you put them on the same plot?}
% \footnote{for the Wage dataset we need some predictive performance measure, i.e, a table  like in the
% simulation--where is it?}

\graphicspath{{/Users/rumpagiri/Documents/NONPARAMETRIC/model_free/papers/REGRESSION}}
\DeclareGraphicsExtensions{.png}

{\begin{figure}[!t]
  \centering
  \includegraphics[width=3.5in, height=2.5in]{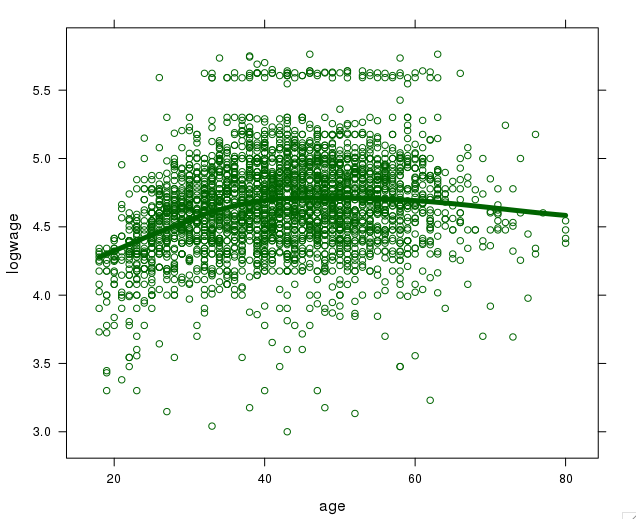}
  \caption{Plot of logwage versus age from Wage dataset (ISLR package)}
  \label{ISLR_data}
\end{figure}}

\graphicspath{{/Users/rumpagiri/Documents/NONPARAMETRIC/model_free/papers/REGRESSION}}
\DeclareGraphicsExtensions{.png}

{\begin{figure}[!t]
  \centering
  \includegraphics[width=3.5in, height=2.5in]{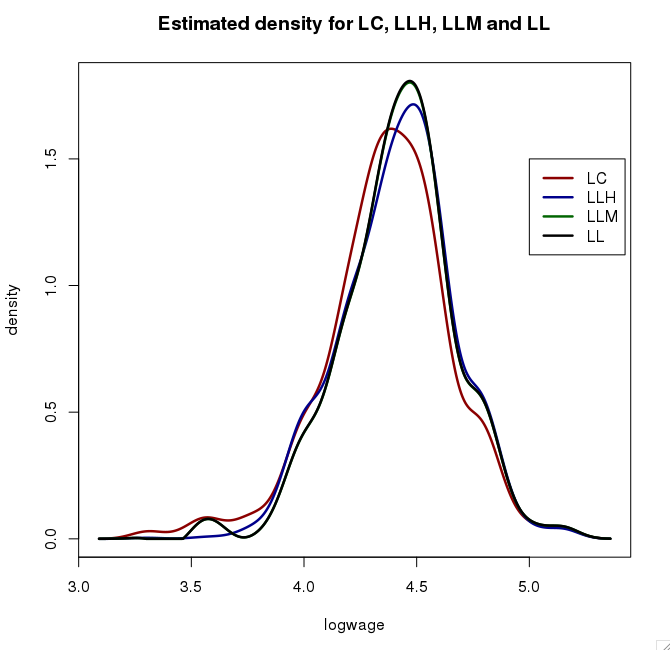}
  \caption{Plot of estimated conditional density function using LC, LLH, LLM and LL methods on ISLR dataset}
  \label{de_islr_density_figure}
\end{figure}}

%\begin{table}
%\centering
%\caption{Point Prediction for ISLR Wage Dataset over Boundary Points with 1-sided CV}
%\label{ISLR_Wage_bp}
%\scalebox{0.7}{
%\begin{tabular}{| w | w | w |}
%\hline
%Method & Bias & MSE  \\
%\hline
%LC & 0.1535513 & 0.109608 \\
%\hline
%LLH & 0.1301558 & 0.10339 \\
%\hline
%LLM & 0.1062107 & 0.097654 \\
%\hline
%\end{tabular}}
%\end{table}
%
%\begin{table}
%\centering
%\caption{Point Prediction for ISLR Wage Dataset over Boundary Points with 1-sided local CV}
%\label{ISLR_Wage_bp}
%\scalebox{0.7}{
%\begin{tabular}{| w | w | w |}
%\hline
%Method & Bias & MSE  \\
%\hline
%LC &  0.05630838 & 0.08854836  \\
%\hline
%LLH &  0.01903547 &  0.08880532 \\
%\hline
%LLM &  0.02055483 &  0.08848614 \\
%\hline
%\end{tabular}}
%\end{table}

\begin{table}
\centering
\caption{Point Prediction for ISLR Wage Dataset}
\label{ISLR_Wage_bp}
\scalebox{0.7}{
\begin{tabular}{| w | w | w |}
\hline
Method & Bias & MSE  \\
\hline
LC &  0.0004954944 &  0.08236025 \\
\hline
LLH &  -0.001962329 &   0.0808793 \\
\hline
LLM &  -6.005305e-05 &  0.08044857 \\
\hline
LL & 0.0002608775 & 0.08055141 \\
\hline
\end{tabular}}
\end{table}

\section {Conclusions}
 
Improved estimation of conditional distributions at boundary points is possible
via local linear smoothing and other methods that, however, do not guarantee that
the resulting estimator is a proper distribution function. In the paper at hand
we propose a simple monotonicity correction procedure that is immediately applicable,
  easy to implement, and performs well with simulated and real data. 

To elaborate, it has been shown using boundary points on simulated datasets that the
LLM distribution estimator outperforms that of LLH and LC as seen by the values of the 
Kolmogorov-Smirnov test statistic, accuracy  of estimated quantiles, and also by its performance in point prediction---the latter finding being entirely unexpected. In contrast, for internal points on
these datasets there seem to be no significant differences between the 3 estimators
using these performance metrics. 

In addition, among all three methods over a wide range of
selected bandwidths the overall best performance is obtained using Monotone Local Linear
Estimation. As can be seen from the point prediction tables, the predictor based on $\bar {F}^{LLM}_{x}(y)$ 
has lower bias compared to $\bar {F}_{x}(y)$ and $\bar {F}^{LLH}_{x}(y)$; this is consistent with the discussion in
Section \ref{LL_est}, i.e. that $\bar {F}^{LLM}_{x}(y)$ has improved performance because of reduced bias in
extrapolation for the boundary case. No such differences in bias are noticed for the case of internal points.
%For the case of estimation which involves regression data distributed non-uniformly the
%LLM method outperforms LLH and LC. 

As in the case  of simulated data, in the real data  example as well  the point prediction
performance of LLM closely matches in performance to that of LL which implies that the LLM distribution estimator can be used for all practical   applications, including point prediction.

\clearpage
\vskip .1in
\noindent 
{\bf Acknowledgements} \\
This research was partially supported by NSF grants DMS 12-23137 and DMS 16-13026.
The authors would like to acknowledge the Pacific Research Platform, NSF Project ACI-1541349 and Larry Smarr (PI, Calit2 at UCSD)
for providing the computing infrastructure used in this project.
 \vskip .1in

%\clearpage

\bibliography{srinjoy_stats}

\begin{thebibliography}{}

\bibitem [\protect \citeauthoryear {%
Fan%
\ \BBA {} Gijbels%
}{%
Fan%
\ \BBA {} Gijbels%
}{%
{\protect \APACyear {1996}}%
}]{%
fan1996local}
\APACinsertmetastar {%
fan1996local}%
\begin{APACrefauthors}%
Fan, J.%
\BCBT {}\ \BBA {} Gijbels, I.%
\end{APACrefauthors}%
\unskip\
\newblock
\APACrefYear{1996}.
\newblock
\APACrefbtitle {Local polynomial modelling and its applications: monographs on
  statistics and applied probability 66} {Local polynomial modelling and its
  applications: monographs on statistics and applied probability 66}\
  (\BVOL~66).
\newblock
\APACaddressPublisher{}{CRC Press, Boca Raton}.
\PrintBackRefs{\CurrentBib}

\bibitem [\protect \citeauthoryear {%
Hall%
, Wolff%
\BCBL {}\ \BBA {} Yao%
}{%
Hall%
\ \protect \BOthers {.}}{%
{\protect \APACyear {1999}}%
}]{%
Hall1999}
\APACinsertmetastar {%
Hall1999}%
\begin{APACrefauthors}%
Hall, P.%
, Wolff, R\BPBI C.%
\BCBL {}\ \BBA {} Yao, Q.%
\end{APACrefauthors}%
\unskip\
\newblock
\APACrefYearMonthDay{1999}{}{}.
\newblock
{\BBOQ}\APACrefatitle {Methods for estimating a conditional distribution
  function} {Methods for estimating a conditional distribution
  function}.{\BBCQ}
\newblock
\APACjournalVolNumPages{Journal of the American Statistical
  Association}{94}{445}{154--163}.
\PrintBackRefs{\CurrentBib}

\bibitem [\protect \citeauthoryear {%
Hansen%
}{%
Hansen%
}{%
{\protect \APACyear {2004}}%
}]{%
hansen2004nonparametric}
\APACinsertmetastar {%
hansen2004nonparametric}%
\begin{APACrefauthors}%
Hansen, B\BPBI E.%
\end{APACrefauthors}%
\unskip\
\newblock
\APACrefYearMonthDay{2004}{}{}.
\newblock
{\BBOQ}\APACrefatitle {Nonparametric estimation of smooth conditional
  distributions} {Nonparametric estimation of smooth conditional
  distributions}.{\BBCQ}
\newblock
\APACjournalVolNumPages{Unpublished paper: Department of Economics, University
  of Wisconsin}{}{}{}.
\PrintBackRefs{\CurrentBib}

\bibitem [\protect \citeauthoryear {%
James%
, Witten%
, Hastie%
\BCBL {}\ \BBA {} Tibshirani%
}{%
James%
\ \protect \BOthers {.}}{%
{\protect \APACyear {2013}}%
}]{%
James2013}
\APACinsertmetastar {%
James2013}%
\begin{APACrefauthors}%
James, G.%
, Witten, D.%
, Hastie, T.%
\BCBL {}\ \BBA {} Tibshirani, R.%
\end{APACrefauthors}%
\unskip\
\newblock
\APACrefYearMonthDay{2013}{}{}.
\newblock
{\BBOQ}\APACrefatitle {ISLR: Data for An Introduction to Statistical Learning
  with Applications in R} {Islr: Data for an introduction to statistical
  learning with applications in r}{\BBCQ}\ [\bibcomputersoftwaremanual].
\newblock
\begin{APACrefURL} \url{http://CRAN.R-project.org/package=ISLR}
  \end{APACrefURL}
\newblock
\APACrefnote{R package version 1.0}
\PrintBackRefs{\CurrentBib}

\bibitem [\protect \citeauthoryear {%
Koenker%
}{%
Koenker%
}{%
{\protect \APACyear {2005}}%
}]{%
koenker2005quantile}
\APACinsertmetastar {%
koenker2005quantile}%
\begin{APACrefauthors}%
Koenker, R.%
\end{APACrefauthors}%
\unskip\
\newblock
\APACrefYear{2005}.
\newblock
\APACrefbtitle {Quantile regression} {Quantile regression}\ (\BNUM~38).
\newblock
\APACaddressPublisher{}{Cambridge University Press, Cambridge}.
\PrintBackRefs{\CurrentBib}

\bibitem [\protect \citeauthoryear {%
Li%
\ \BBA {} Racine%
}{%
Li%
\ \BBA {} Racine%
}{%
{\protect \APACyear {2007}}%
}]{%
li2007nonparametric}
\APACinsertmetastar {%
li2007nonparametric}%
\begin{APACrefauthors}%
Li, Q.%
\BCBT {}\ \BBA {} Racine, J\BPBI S.%
\end{APACrefauthors}%
\unskip\
\newblock
\APACrefYear{2007}.
\newblock
\APACrefbtitle {Nonparametric econometrics: theory and practice} {Nonparametric
  econometrics: theory and practice}.
\newblock
\APACaddressPublisher{}{Princeton University Press, Princeton}.
\PrintBackRefs{\CurrentBib}

\bibitem [\protect \citeauthoryear {%
Politis%
}{%
Politis%
}{%
{\protect \APACyear {2013}}%
}]{%
Politis2013}
\APACinsertmetastar {%
Politis2013}%
\begin{APACrefauthors}%
Politis, D\BPBI N.%
\end{APACrefauthors}%
\unskip\
\newblock
\APACrefYearMonthDay{2013}{}{}.
\newblock
{\BBOQ}\APACrefatitle {Model-free model-fitting and predictive distributions}
  {Model-free model-fitting and predictive distributions}.{\BBCQ}
\newblock
\APACjournalVolNumPages{Test}{22}{2}{183--221}.
\PrintBackRefs{\CurrentBib}

\bibitem [\protect \citeauthoryear {%
Politis%
}{%
Politis%
}{%
{\protect \APACyear {2015}}%
}]{%
politis2015model}
\APACinsertmetastar {%
politis2015model}%
\begin{APACrefauthors}%
Politis, D\BPBI N.%
\end{APACrefauthors}%
\unskip\
\newblock
\APACrefYear{2015}.
\newblock
\APACrefbtitle {Model-Free Prediction and Regression} {Model-free prediction
  and regression}.
\newblock
\APACaddressPublisher{}{Springer, New York}.
\PrintBackRefs{\CurrentBib}

\bibitem [\protect \citeauthoryear {%
Schucany%
}{%
Schucany%
}{%
{\protect \APACyear {2004}}%
}]{%
schucany2004kernel}
\APACinsertmetastar {%
schucany2004kernel}%
\begin{APACrefauthors}%
Schucany, W\BPBI R.%
\end{APACrefauthors}%
\unskip\
\newblock
\APACrefYearMonthDay{2004}{}{}.
\newblock
{\BBOQ}\APACrefatitle {Kernel smoothers: an overview of curve estimators for
  the first graduate course in nonparametric statistics} {Kernel smoothers: an
  overview of curve estimators for the first graduate course in nonparametric
  statistics}.{\BBCQ}
\newblock
\APACjournalVolNumPages{Statistical Science}{}{}{663--675}.
\PrintBackRefs{\CurrentBib}

\bibitem [\protect \citeauthoryear {%
Wand%
\ \BBA {} Jones%
}{%
Wand%
\ \BBA {} Jones%
}{%
{\protect \APACyear {1994}}%
}]{%
wand1994kernel}
\APACinsertmetastar {%
wand1994kernel}%
\begin{APACrefauthors}%
Wand, M\BPBI P.%
\BCBT {}\ \BBA {} Jones, M\BPBI C.%
\end{APACrefauthors}%
\unskip\
\newblock
\APACrefYear{1994}.
\newblock
\APACrefbtitle {Kernel smoothing} {Kernel smoothing}.
\newblock
\APACaddressPublisher{}{CRC Press, Boca Raton}.
\PrintBackRefs{\CurrentBib}

\bibitem [\protect \citeauthoryear {%
Yu%
\ \BBA {} Jones%
}{%
Yu%
\ \BBA {} Jones%
}{%
{\protect \APACyear {1998}}%
}]{%
yu1998local}
\APACinsertmetastar {%
yu1998local}%
\begin{APACrefauthors}%
Yu, K.%
\BCBT {}\ \BBA {} Jones, M.%
\end{APACrefauthors}%
\unskip\
\newblock
\APACrefYearMonthDay{1998}{}{}.
\newblock
{\BBOQ}\APACrefatitle {Local linear quantile regression} {Local linear quantile
  regression}.{\BBCQ}
\newblock
\APACjournalVolNumPages{Journal of the American statistical
  Association}{93}{441}{228--237}.
\PrintBackRefs{\CurrentBib}

\bibitem [\protect \citeauthoryear {%
Yu%
, Lu%
\BCBL {}\ \BBA {} Stander%
}{%
Yu%
\ \protect \BOthers {.}}{%
{\protect \APACyear {2003}}%
}]{%
yu2003quantile}
\APACinsertmetastar {%
yu2003quantile}%
\begin{APACrefauthors}%
Yu, K.%
, Lu, Z.%
\BCBL {}\ \BBA {} Stander, J.%
\end{APACrefauthors}%
\unskip\
\newblock
\APACrefYearMonthDay{2003}{}{}.
\newblock
{\BBOQ}\APACrefatitle {Quantile regression: applications and current research
  areas} {Quantile regression: applications and current research areas}.{\BBCQ}
\newblock
\APACjournalVolNumPages{Journal of the Royal Statistical Society: Series D (The
  Statistician)}{52}{3}{331--350}.
\PrintBackRefs{\CurrentBib}

\end{thebibliography}
\bibliographystyle{apacite}

\end{document}